\documentclass[aps,pra,twocolumn,amsmath,amssymb,showpacs]{
revtex4-1}
\usepackage[latin9]{inputenc}
\usepackage{amstext}
\usepackage{amssymb}
\usepackage{graphicx}
\usepackage{color}

\usepackage{graphicx}
\usepackage{dcolumn}
\usepackage{physics}
\usepackage{hyperref}
\usepackage{bm}
\usepackage{textcomp}

\usepackage{amsmath,amsthm}
\usepackage[scriptsize]{subfigure} 
\usepackage{array}
\usepackage{floatrow}
\usepackage{epstopdf}

\usepackage[normalem]{ulem}

\makeatletter

\newfloatcommand{capbtabbox}{table}[\captop][0.8\FBwidth]

\makeatother

\begin{document}


\title{ Diffusion Monte Carlo methods for Spin-Orbit-Coupled ultracold Bose gases
}

\author{J. S\'anchez-Baena}\email{juan.sanchez.baena@upc.edu}
 \author{J. Boronat}\email{jordi.boronat@upc.edu}%
\author{F. Mazzanti}\email{ferran.mazzanti@upc.edu}
\affiliation{%
 Departament de F\'isica, Universitat Polit\`ecnica de 
Catalunya,
Campus Nord B4-B5, E-08034, Barcelona, Spain\\
}%


\date{\today}

\begin{abstract}

We present two Diffusion Monte Carlo (DMC) algorithms for systems of
ultracold quantum gases featuring synthetic spin-orbit interactions.
The first one is a spin-integrated DMC method which provides fixed-phase energy estimates. The second one is a discrete spin generalisation of the T-moves
spin-orbit DMC~\cite{mitas}
, which provides an upper bound to the fixed-phase
energy. The former is a more
accurate method but it is restricted to spin-independent two-body
interactions.
We report a comparison between both algorithms for
different systems. As a check of the efficiency of both methods, we
compare the DMC energies with results obtained with other numerical
methods, finding agreement between both estimations.

\end{abstract}

\maketitle


\section{\label{sec:introduction}Introduction}

The interplay between the electron spin and its momentum, known as
spin-orbit coupling (SOC), is an effect of major relevance when
studying a wide variety of systems in the field of solid-state
physics, such as Majorana fermions~\cite{majorama}, spintronic
devices~\cite{spintronic} or topological
insulators~\cite{insulator}. The realization in the last few years of
a synthetic SOC interaction in ultracold atomic gases, by exploiting
the space-dependent coupling of the atoms with a properly designed
configuration of laser
beams~\cite{spielman},~\cite{dalibard},~\cite{goldman},~\cite{zhang}, represents an
important achievement. More interestingly, these new realizations allow for a
better understanding of the effects induced by the SOC interaction, since
ultracold quantum gases are highly controllable and
tunable~\cite{stringari}.  Ultracold SOC quantum gases have been
studied in the dilute regime~\cite{stringari2}, showing the rise of
new exotic phases, such as a spin-polarized plane wave phase and a
stripe phase. This stripe phase has been recently observed by Li
\textit{et al.}~\cite{li} showing specific properties of a supersolid
phase.

Up to now, the theoretical approaches used in the study of SOC gases
rely on the mean-field approximation. This theory is expected to be
valid when the gas parameter is very small, $na^3 \leq 10^{-5}$, but
beyond this limit one is faced with beyond-mean field terms. A way of
surpassing the range of applicability of the mean-field approximation
is the use of quantum Monte Carlo (QMC) methods, which are not based
on any perturbative scheme.  In the present work, we use QMC to study
these ultracold atomic gases featuring a synthetic SOC interaction. In
particular, we work with the Diffusion Monte Carlo (DMC) method, which
is a stochastic method intended for solving the imaginary-time
many-body Schr\"odinger equation. The action of the imaginary-time
propagator $\exp [ -\tau \hat{H} ]$ is implemented as a set of
transformations to a list of points in coordinate space (commonly
called \textit{walkers}) that represent statistically the wave
function.  In the limit $\tau \rightarrow \infty$, the ground state
dominates while excited-state contributions are exponentially damped,
providing an exact estimate of the ground-state energy and of any
observable commuting with $\hat{H}$. If the ground state of the system
of interest is complex (which is the case when the SOC term is
present), it is necessary to invoke the fixed-phase approximation (FPA),
which provides an upper bound to the ground-state energy.

Previous DMC calculations with SOC terms in the Hamiltonian have been
carried out in the study of electronic
structures~\cite{mitas,mitas2}, quantum dots in
semi-conductors~\cite{pederiva2}, and repulsive Fermi
gases~\cite{pederivamitas}. A DMC method incorporating the SOC terms
that arise in electronic systems has already been
developed~\cite{mitas}. In this method, the authors implement
the spin-orbit term of the propagator through the use of the T-moves
technique~\cite{tmoves}. They also use a regularized, continuous
representation of the spin degrees of freedom. In order to control the sign problem that the SOC
terms introduce in the propagator, the authors of Ref. \cite{mitas} define an
effective Hamiltonian in such a way that the propagator becomes
positive-definite.  It can be shown that the estimations obtained with
this effective Hamiltonian yield an upper bound to the
fixed-phase energy~\cite{ceperley}. In the present
paper, we adapt the T-moves DMC algorithm of Ref.~\cite{mitas} to the usual,
discrete representation of the spin, and show how to treat the
synthetic SOC present in ultracold quantum gases. We also introduce a
different method for treating the SOC terms of the propagator, loosely
based on Ref.~\cite{pederiva}, which consists on propagating the wave
function integrated over all spin configurations. In doing so, we
avoid almost completely the sign problem induced by SOC terms, meaning
that no effective Hamiltonian needs to be defined.

This paper is organized as follows. In Sec. \ref{sec:Hamiltonian}, we
discuss the form of the Hamiltonian as well as several kinds of
spin-orbit couplings of interest in the field of cold Bose gases.  The
reduced units used in this work are introduced in
Sec.~\ref{sec:units}.  In Sec. \ref{sec:method1} we present the
details concerning the Spin-integrated DMC. We derive the fixed-phase,
spin-orbit propagator to first order and elaborate on its implementation within the
DMC framework. We also discuss how to estimate the energy and provide
a scheme of the DMC algorithm. In Sec. \ref{sec:tmoves}, we show how
to implement discrete spin sampling within the T-moves DMC , as well
as how to implement the SOC terms introduced in
Sec. \ref{sec:Hamiltonian}. In Sec. \ref{sec:results}, we compare both
DMC methods in one and two-body problems (\ref{sec:fewbody}) and in some many-body
cases (\ref{sec:manybody}).  Finally, in Sec. \ref{sec:conclusions}, we
summarize the main conclusions of our work.

\section{\label{sec:Hamiltonian}Hamiltonian}

The system studied in this work is formed by an ultracold gas of $N$
bosons of mass $M$ with pseudo-spin $1/2$ under the effect of
synthetic spin-orbit coupling~\cite{stringari}. The generic form of
the Hamiltonian is:
\[
\hat{H} = \sum_{k=1}^N \left[ \frac{\hat P_k^2}{2M} +
  \hat{V}_{k}^{\text{1b}} + \hat{W}_k^{\text{SOC}} \right] +
\hat{V}^{\text{2b}} \ ,
\label{Hamiltonian_a1}
\]
with $\hat{V}_{k}^{\text{1b}}$ and $\hat{V}^{\text{2b}}$ momentum
independent, local, one- and two-body interactions,
respectively. Notice that $\hat V^{\rm 2b}$ can depend on the spin
configuration. In much the same way, $\hat{W}_k^{\text{SOC}}$ stands
for a one-body, momentum and spin-dependent potential. The
ones considered in this work are the Rashba, Weyl and Raman
interactions given by 
\begin{align}
\hat{W}_k^{\text{Rs}} &= \frac{\lambda_{\text{Rs}} \hbar}{2}
\left[ \hat{P}_k^y \hat{\sigma}_k^x - \hat{P}_k^x \hat{\sigma}_k^y
  \right] \label{Wrashba} \\
\hat{W}_k^{\text{Rm}} &=
\frac{\lambda_{\text{Rm}} \hbar}{M} \hat{P}_k^x \hat{\sigma}_k^z +
\frac{\lambda_{\text{Rm}}^2 \hbar^2}{2M} - \frac{\Omega}{2}
\hat{\sigma}_k^x \label{Wraman} \\
\hat{W}_k^{\text{We}} &=
\frac{\lambda_{\text{We}} \hbar}{M} \left[ \hat{P}_k^x
  \hat{\sigma}_k^x + \hat{P}_k^y \hat{\sigma}_k^y + \hat{P}_k^z
  \hat{\sigma}_k^z \right] + \frac{\lambda_{\text{We}}^2
  \hbar^2}{2M}
\label{Wweyl} \ ,
\end{align}
with $\hat{P}_k^{\alpha}$ the $\alpha$-component of the momentum
operator of particle $k$, $\hat{\sigma}_k^{x,y,z}$ the Pauli matrices
associated to particle $k$, $\Omega$ the Rabi frequency, and
$\lambda_\alpha$ ($\alpha=\{$ Rs, We, Rm$\}$)
the strength of the corresponding SOC interaction. The general form of the
two-body potential is:
\begin{equation}
  \hat V^{\rm 2b} =
  \sum_{k<l} \left[
    \sum_{s_k,s_l} V^{\rm 2b}_{s_k,s_l}(r_{kl}) | s_k, s_l \rangle
    \langle s_k, s_l |
    \right] \ ,
  \label{V2b_a1}
\end{equation}
where $s_k, s_l$ assign values $\pm 1$ to the $z$-component of the
spin of particles $k$ and $l$, while $V^{\rm 2b}_{s_k,s_l}(r_{kl})$
is a central, short-ranged potential that can be different for the
different channels corresponding to $s_k$ and $s_l$. 
In the numerical examples of Sec. \ref{sec:results} we use a soft-core force, defined by: 
\begin{equation}
  V_{s_k,s_l}(r) = V_0(s_k,s_l) \,\theta\!\left( R_0(s_k,s_l)
  -r\right)) \ .
\label{V2b}
\end{equation}
If the two-body interaction is taken to be spin-independent,
$V_0(s_k,s_l) = V_0$ and $R_0(s_k,s_l) = R_0$.

The one-body potential used in some of the calculations below is:
\begin{equation}
\hat{V}^{\text{1b}} = \frac{1}{2} M \omega^2 ( \hat{X}^2 + \hat{Y}^2
+ \hat{Z}^2 ) \ .
\label{V1b}
\end{equation}
\subsection{\label{sec:units}Reduced units for the different kinds of SOC interactions}

Due to the different spin dependence, we use different length and
energy scales in each case. These are the following: for the Rashba
interaction, we set the length and energy units to
\begin{equation}
a_{\text{Rs}} = \frac{1}{\lambda_{\text{Rs}} M}
\,\,\, , \,\,\,
e_{\text{Rs}} = \frac{ \hbar^2 }{ 2 M a_{\text{Rs}}^2 } =
\frac{\hbar^2 \lambda_{\text{Rs}}^2 M}{2} \ ,
\label{Rashba_units}
\end{equation}
while for the Raman interaction
\begin{equation}
a_{\text{Rm}} = \frac{ \eta_{\text{Rm}} }{ \lambda_{\text{Rm}} }
\,\,\, , \,\,\,
e_{\text{Rm}} = \frac{ \hbar^2 }{ 2 M a_{\text{Rm}}^2 } = \frac{
  \hbar^2 \lambda_{\text{Rm}}^2 }{ 2 M \eta_{\text{Rm}}^2 } \ .
\label{Raman_units}
\end{equation}
with $\eta_{Rm}$ a dimensionless scaling factor that we vary depending
on the density. Finally, for the Weyl Hamiltonian we use
\begin{equation}
a_{\text{We}} = \frac{ \eta_{\text{We}} }{ 2 \lambda_{\text{We}} }
\,\,\, , \,\,\,
e_{\text{We}} = \frac{ \hbar^2 }{ 2 M a_{\text{We}}^2 } = \frac{ 2
  \hbar^2 \lambda_{\text{We}}^2 }{ M \eta_{\text{We}}^2 } \ .
\label{Weyl_units}
\end{equation}

In terms of these, the interactions read
\begin{align}
  \hat{W}_k^{\text{Rashba}} &= \left[ \hat{P}_k^y \hat{\sigma}_k^x -
    \hat{P}_k^x \hat{\sigma}_k^y \right] \label{Wrashba_red} \\
\hat{W}_k^{\text{Raman}} &= 2 \eta_{\text{Rm}} \hat{P}_k^x
\hat{\sigma}_k^z + \eta_{\text{Rm}}^2 - \frac{\Omega}{2}
\hat{\sigma}_k^x \label{Wraman_red} \\ 
\hat{W}_k^{\text{Weyl}}  &= \eta_{\text{We}} \left[ \hat{P}_k^x
  \hat{\sigma}_k^x + \hat{P}_k^y \hat{\sigma}_k^y + \hat{P}_k^z
  \hat{\sigma}_k^z \right] +
\frac{\eta_{\text{We}}^2}{4} \label{Wweyl_red} 
\end{align}
where all quantities are dimensionless. The same applies to the
soft-core potential and harmonic trap of Eqs.~(\ref{V2b})
and~(\ref{V1b}).

\section{\label{sec:method1}The Spin-integrated DMC (SIDMC) method}

\subsection{\label{sec:method1A}
  The spin-orbit propagator in the fixed-phase approximation}

In this section we derive a suitable form of the propagator required
to simulate spin-orbit problems, under the assumption that the
two-body interaction $\hat{V}^{\text{2b}}$ is spin-independent.
The imaginary time evolution of state $\ket{\Psi(\tau)}$ is given by
\begin{equation}
 \ket{\Psi(\tau + \Delta \tau)} = \exp \left[ -\Delta \tau \hat{H}
   \right] \ket{\Psi(\tau)} \ .
 \label{firstev}
\end{equation}
Projecting on $\bra*{\vec{R}',\vec{S}'}$ and introducing an identity,
Eq. \ref{firstev} can be written as:
\begin{eqnarray}
  \psi(\vec{R}',\vec{S}',\tau + \Delta \tau) & = & 
  \sum_{\vec{S}} \int \vec{dR} \bra*{\vec{R}',\vec{S}'} \exp \left[
    -\Delta \tau \hat{H} \right] \!\! \ket*{\vec{R},\vec{S}} \nonumber \\  & \times &  \psi(\vec{R},\vec{S},\tau) \ ,
\label{secondev}
\end{eqnarray}
where $\vec{R}$ and $\vec{S}$ stand for the position and spin
coordinates of the $N$ particles. For the sake of clarity, we also
define
\begin{align}
 \hat{H}_0 &= \sum_{k=1}^N \left[ \frac{P_k^2}{2M} + \hat{V}_{k}^{\text{1b}} +\sum_{l<k}^N \hat{V}_{k,l}^{\text{2b}} \right] \\
 \hat{W} &= \sum_{k=1}^N \hat{W}_k^{\text{SOC}} \ .
\end{align}
Up to $\order{\Delta \tau}$, Eq.~(\ref{secondev}) can be written as:
\begin{eqnarray}
  \psi(\vec{R}' & , & \vec{S}',\tau + \Delta \tau) = \sum_{\vec{S}}
  \int \vec{dR}  \int \vec{dR''} \nonumber \\
   & & \times \bra*{\vec{R}',\vec{S}'} \exp \left[ -\Delta \tau \hat{W}
     \right] \ket*{\vec{R}'',\vec{S}} \\
    & & \cross \bra*{\vec{R}''} \exp \left[ -\Delta \tau
    \hat{H}_0 
    \right] \ket*{\vec{R}} \psi(\vec{R},\vec{S},\tau) + \order{\Delta
   \tau^2} \ ,\nonumber
\end{eqnarray}
where the term corresponding to $\hat H_0$ in the splitting of $\hat
H$ in the propagator is spin-independent.

In this way, the propagator reads:
\begin{eqnarray}
  G( \vec{R},\vec{S} \rightarrow \vec{R}',\vec{S}') & = & \int
  \vec{dR''} \bra*{\vec{R}',\vec{S}'} \exp \left[ -\Delta \tau \hat{W}
    \right] \ket*{\vec{R}'',\vec{S}} \nonumber \\
  & & \times
  \bra*{\vec{R}''} \exp \left[ -\Delta \tau \hat{H}_0 \right]
  \ket*{\vec{R}} \ .
 \label{propagator}
\end{eqnarray}
This propagator can have complex contributions coming from the Pauli
matrices appearing in the spin-orbit interaction, and therefore
sampling it is not possible. In order to bypass this problem, we
resort to the fixed-phase approximation~\cite{mitas} where all
quantities involved are real.

Knowing the general expression of the propagator written above, we can
deduce its reduction to the fixed-phase approximation.  This can be
done comparing the imaginary-time Schr\"odinger equation for the wave
function and for its magnitude, which is the main quantity of interest
in the FPA.
For the full wave function, one has
\begin{eqnarray}
-\pdv{\psi ( \vec{R},\vec{S} ) }{\tau} & = & \left[ \sum_{k=1}^N
  \left( -\frac{\hbar^2}{2 M} \nabla_k^2 + V^{\text{1b}}_k
  (\vec{r}_k) + \sum_{l<k}^N \hat{V}_{k,l}^{\text{2b}}(r_{kl})
  \right) \right] \nonumber \\
& & \times \psi(\vec{R},\vec{S},\tau)
\label{propagator_full} \\
& &  + \sum_{\vec{S}'} \int \vec{dR}'
\bra*{\vec{R},\vec{S}}\hat{W}_{\text{spin}} \ket*{\vec{R}',\vec{S}'}
\psi(\vec{R}',\vec{S}',\tau) \ ,\nonumber
\end{eqnarray}
while for the magnitude $\rho(\vec{R},\vec{S})$ of
$\psi(\vec{R},\vec{S})$ the equation reads
\begin{eqnarray}
 &-&\pdv{ \rho ( \vec{R},\vec{S} ) }{\tau} = \left[ \sum_{k=1}^N
    \left( -\frac{\hbar^2}{2 M} \nabla_k^2 +\frac{\hbar^2}{2 M} \abs*{
      \vec{\nabla}_k \Phi(\vec{R},\vec{S},\tau) }^2
    \right. \right. \nonumber \\    
& & \left. \left. + V^{\text{1b}}_k (\vec{r}_k) + \sum_{l<k}^N
    \hat{V}_{k,l}^{\text{2b}}(\vec{r}_k, \vec{r}_l) \right) \right]
  \rho(\vec{R},\vec{S},\tau) \nonumber \\
& & + \sum_{\vec{S}'} \int \vec{dR}'
  \bra*{\vec{R},\vec{S}}\hat{w}_{\text{Re}} \ket*{\vec{R}',\vec{S}'}
  \rho(\vec{R}',\vec{S}',\tau) \ ,
\label{propagator_fixedphase}
\end{eqnarray}
where
\begin{equation}
\psi ( \vec{R},\vec{S},\tau ) = \rho ( \vec{R},\vec{S},\tau
)\exp\left[ i\Phi(\vec{R},\vec{S},\tau) \right]
\end{equation}
and
\begin{equation}
 \bra*{\vec{R},\vec{S}}\hat{w}_{\text{Re}} \ket*{\vec{R}',\vec{S}'} =
 \text{Re} \left\{ \!\!\bra*{\vec{R},\vec{S}}\hat{W}_{\text{spin}}
 \ket*{\vec{R}',\vec{S}'} \frac{ e^{i \Phi(\vec{R}',\vec{S}',\tau)} }{
   e^{ i\Phi(\vec{R},\vec{S},\tau) } } \right\} \ .
 \label{w_RE_guais}
\end{equation}
In the FPA, $\Phi(\vec{R},\vec{S},\tau)$ is
independent of $\tau$ and $\hat V_\Phi= \sum_{k=1}^N |\nabla_k \Phi(\vec{R},\vec{S},\tau)|^2$
becomes a local interaction in positions and
spins. Equations~(\ref{propagator_full})
and~(\ref{propagator_fixedphase}) have a similar structure, and thus
comparing terms in each, we can get the FPA form of
the propagator in Eq.~(\ref{propagator}):
\begin{eqnarray}
  G_{\text{FP}} \!\! & ( & \!\! \vec{R},\vec{S} \rightarrow \vec{R}',\vec{S}') =
  \bra*{\vec{R}',\vec{S}'} \exp \left[ -\Delta \tau
    \hat{H}^{\text{FP}} \right] \ket*{\vec{R},\vec{S}} \nonumber \\
  & = & \int \vec{dR''}
  \bra*{\vec{R}',\vec{S}'} \exp \left[ -\Delta
    \tau \left( \hat{w}_{\text{Re}} + \hat{V}_{\Phi} \right) \right]
  \ket*{\vec{R}'',\vec{S}}
  \nonumber \\
  & & \times \bra*{\vec{R}''} \exp \left[ -\Delta \tau \hat{H}_0
    \right] \ket*{\vec{R}} + \order{\Delta \tau^2} \ ,
  \label{propagator_FP}
\end{eqnarray}
with
\begin{equation}
\hat{H}^{\text{FP}} = \hat{H}_0 + \hat{w}_{\text{Re}} + \hat{V}_{\Phi}
\label{HFP}
\end{equation}
the fixed-phase Hamiltonian. In the FPA one has
to impose a certain form for the phase. In this work we impose it to be
the sum of one-body terms
\begin{equation}
 \Phi(\vec{R},\vec{S}) = \sum_{k=1}^N \phi_k(\vec{r}_k,s_k) \ .
 \label{trialphase}
\end{equation}
Due to the form of the spin-orbit potential, we can evaluate the
integral in Eq.~\ref{propagator_FP}. For the Raman SOC of
Eq.~\ref{Wraman}, the matrix element of the spin-dependent part of the
potential is

\begin{eqnarray}
  \langle \vec{R}',\vec{S}' \!\! & | & \!\!
  \hat{w}_{\text{Re}} + \hat{V}_{\Phi}
  \ket*{\vec{R}'',\vec{S}} = \sum_{k=1}^N \left[ \prod_{l\neq k}^N
    \delta_{\vec{r}'_l,\vec{r}''_l} \delta_{s'_l,s_l} \right]
  \nonumber \\  
& \times &  \left[ \frac{\lambda \hbar}{M} \delta_{y_k' , y_k''}
    \delta_{z_k',z_k''} \frac{d \delta_{x_k',x_k''}}{d x_k'}
    \bra*{s_k'} \hat{\sigma}_k^z \ket*{s_k} \sin ( \Delta \phi_k )
    \right. \nonumber \\  
& & + \left. \frac{\lambda^2}{2M}
    \delta_{\vec{r}_k',\vec{r}_k''} \delta_{s_k',s_k} -
    \frac{\Omega}{2} \bra*{s_k'} \hat{\sigma}_k^x \ket*{s_k} \cos (
    \Delta \phi_k ) \right. \nonumber \\ [2mm]
& & \left. + \,\,
    \abs*{\vec{\nabla}_k \phi_k}^2 \delta_{\vec{r}_k',\vec{r}_k''}
    \delta_{s_k',s_k} \right] \ ,
\end{eqnarray}
where
\begin{equation}
\Delta \phi_k = \phi_k(\vec{r}_k'',s_k) - \phi_k(\vec{r}_k',s_k') \ .
\end{equation}
Since the spinless part of the propagator is given by~\cite{Guardiola}, one has
\begin{eqnarray}
\langle \vec{R}'' & | & \exp \left[ -\Delta \tau \hat{H}_0
    \right] \ket*{\vec{R}} = \exp \left[ -\frac{M}{2\hbar^2 \Delta
      \tau} \left( \vec{R}'' - \vec{R} \right)^2 \right] \nonumber \\
& & \times \exp \left[ \Delta \tau \left( E_s -
     \frac{V_0(\vec{R}'') + V_0(\vec{R}'')}{2} \right) \right]
\end{eqnarray}
with $V_0$ the spinless part of the potential entering in $\hat H_0$ 
and $E_s$ the common energy shift used in the DMC algorithm.
Up to $\order{\Delta \tau}$, the integral in
Eq.~(\ref{propagator_FP}) yields
{\small
\begin{align}
 &G_{\text{FP}} ( \vec{R},\vec{S} \rightarrow \vec{R}',\vec{S}')
  = \bra*{\vec{R}'} \exp \left[ -\Delta \tau \hat{H}_0 \right]
  \ket*{\vec{R}} \nonumber \\ &\cross \left\{
  \delta_{\vec{S}',\vec{S}} - \Delta \tau \sum_{k=1}^N \left[
    \prod_{l\neq k}^N \delta_{s'_l,s_l} \right] \eval{ \left[ \frac{\lambda
      \hbar}{M} \bra*{s_k'} \hat{\sigma}_k^z \ket*{s_k} \cos ( \Delta
    \phi_k ) \right. \right. \right. \nonumber
    \\ &\left. \left. \left. \cross \pdv{\phi_k}{x_k''} + \frac{\lambda^2}{2M} \delta_{s_k',s_k} -
    \frac{\Omega}{2} \bra*{s_k'} \hat{\sigma}_k^x \ket*{s_k} \cos (
    \Delta \phi_k ) \right. \right. \right. \nonumber \\
    &\left. \left. \left. + \abs*{\vec{\nabla}_k \phi_k}^2 \delta_{s_k',s_k}
    \right] }_{ \vec{R}'' = \vec{R}' } \right\} \ .
\label{propagator2}
\end{align}
}%
For the Rashba and Weyl SOC interactions, a similar procedure has to
be carried out. However, one has to expand the element
$  \bra*{\vec{R}',\vec{S}'} \exp \left[ -\Delta
    \tau \left( \hat{w}_{\text{Re}} + \hat{V}_{\Phi} \right) \right]
  \ket*{\vec{R}'',\vec{S}}$ in Eq.~(\ref{propagator_FP})
up to order $\Delta \tau^2$. This is because 
the terms originated from the matrix
element of $\hat{w}_{\text{Re}}$ are proportional to $\xi_k=r_k' - r_k$, and
thus, the elements arising from $\hat{w}^2_{\text{Re}}$ generate
contributions of order $\xi_k^2$ and $\xi_k \xi_l$.
Since $\xi_k$ represents the displacement of particle $k$ due to the
standard DMC Gauss-Drift-Branching (GDB) process, this quantity is of
$\order{\sqrt{\Delta \tau}}$. However, in
the numerical experiments conducted, we have not found a significant
impact on the results when these terms are dropped.

Following with the derivation of the propagator in
Eq.~(\ref{propagator2}), we define a new operator $\hat{O}$ as
\begin{equation}
  \bra*{\vec{S}'} \hat{O} (\vec{R}') \ket*{\vec{S}} =
       {
         G_{\text{FP}} ( \vec{R},\vec{S} \rightarrow
         \vec{R}',\vec{S}')
         \over
         \bra*{\vec{R}'} \exp \left[ -\Delta \tau \hat{H}_0 \right]
         \ket*{\vec{R}}
       } \ .
\label{propagator3}
\end{equation}
while, up to $\order{\Delta \tau}$, Eq.~(\ref{propagator2}) can be rewritten as:
\begin{align}
 &G_{\text{FP}} ( \vec{R},\vec{S} \rightarrow \vec{R}',\vec{S}') \simeq \bra*{\vec{R}'} \exp \left[ -\Delta \tau \hat{H}_0 \right] \ket*{\vec{R}} \nonumber \\
 &\cross \prod_{k=1}^N \left\{ \delta_{s_k',s_k} - \Delta \tau \eval{ \left[ \frac{\lambda \hbar}{M} \bra*{s_k'} \hat{\sigma}_k^z \ket*{s_k} \cos ( \Delta \phi_k ) \right. \right. \right. \nonumber \\
 &\left. \left. \left. \times \pdv{\phi_k}{x_k''} + \frac{\lambda^2}{2M} \delta_{s_k',s_k} - \frac{\Omega}{2} \bra*{s_k'} \hat{\sigma}_k^x \ket*{s_k} \cos ( \Delta \phi_k ) \right. \right. \right. \nonumber \\
 &\left. \left. \left. + \abs*{\vec{\nabla}_k \phi_k}^2  \delta_{s_k',s_k} \right] }_{\vec{r}_k'' = \vec{r}_k'} \right\} \nonumber \\
 &= \bra*{\vec{R}'} \exp \left[ -\Delta \tau \hat{H}_0 \right] \ket*{\vec{R}} \prod_{k=1}^N \bra*{s_k'} \hat{O}_k (\vec{r}_k') \ket*{s_k}
\label{propagator4}
\end{align}
where we have used the approximation $(1-\Delta t\sum x_i) \approx
\prod(1-\Delta t x_i)$ which is exact to order $\Delta t$.
In this way, the matrix element of the new operator $\hat
O$ becomes the product of matrix elements of single-particle operators
$\hat{O}_k$, as shown in the expression above. 

Note that, for the Rashba and Weyl SOCs, the matrix elements
$\bra*{s_k'} \hat{O}_k \ket*{s_k}$ depend both on $\vec{r}_k'$ and
$\vec{r}_k$. For the sake of simplicity, in the following we omit the
$r_k$ and $r'_k$ labels. The imaginary time evolution equation
for the magnitude of the wave function, within
the fixed-phase approximation and to order $\Delta t$, becomes
\begin{eqnarray}
  \rho ( \vec{R}',\vec{S}',\tau + \Delta \tau ) & = & \sum_{\vec{S}}
  \int \vec{dR} \left\{ \prod_{k=1}^N \bra*{s_k'} \hat{O}_k \ket*{s_k}
  \right.
  \label{iterho_a1} \\
  & \times &  \bra*{\vec{R}'} \exp \left[ -\Delta \tau \hat{H}_0
    \right] \ket*{\vec{R}} \rho ( \vec{R},\vec{S},\tau ) \Bigg\} \ .
  \nonumber 
\end{eqnarray}
However, in DMC simulations the object that is propagated is 
$f(\vec R, \vec S,\tau) = \rho ( \vec{R},\vec{S},\tau )\rho_T ( \vec{R},\vec{S} )$,
with $\rho_T ( \vec{R},\vec{S} )$ the magnitude of a given importance sampling
trial function. From Eq.~(\ref{iterho_a1}) one readily sees that
\begin{eqnarray}
 f( \vec{R}',\vec{S}',\tau & + & \Delta \tau ) = \sum_{\vec{S}} \int
 \vec{dR} \left\{ \prod_{k=1}^N \bra*{s_k'} \hat{O}_k \ket*{s_k} \right. \label{iterhois} \\
 & \times & \left. \bra*{\vec{R}'} \exp \left[ -\Delta \tau \hat{H}_0
   \right] \ket*{\vec{R}} \frac{\rho_T ( \vec{R}',\vec{S}' ) }{\rho_T
   ( \vec{R},\vec{S} ) } f ( \vec{R},\vec{S},\tau ) \right\} \nonumber
\end{eqnarray}
In order to implement this equation, we need the propagator to be
positive-definite. However, due to the spin-orbit coupling, the matrix
elements of the propagator do not fulfill this condition. Despite
this, if we propagate the spin-integrated
form of the magnitude of the importance sampling function $f$ of Eq.~(\ref{iterhois}),
this problem is greatly reduced. Therefore, we propagate the quantity
\begin{equation}
 F(\vec{R},\tau) = \sum_{\vec{S}} f ( \vec{R},\vec{S},\tau ) \ .
\label{intrho}
\end{equation}
In order to progress, we impose the magnitude of the trial wave
function to be spin-independent i.e. $\rho_T(\vec{R},\vec{S}) =
\rho_T(\vec{R})$. After $j$ time steps, one gets
\begin{eqnarray}
  F ( \vec{R}^{(j)},j\Delta \tau ) & = &
  \sum_{\vec{S}^{(j)},...,\vec{S}^{(0)}} \int \vec{dR}^{(j-1)}
  \cdot\cdot\cdot \vec{dR}^{(0)} \nonumber \\
  & \times & \prod_{n=1}^j \left( \prod_{k=1}^N \bra*{s^{(n)}_k} \hat{O}_k
    \ket*{s^{(n-1)}_k} \right) \nonumber \\
  & \times & \prod_{n=1}^j \bra*{\vec{R}^{(n)}} \exp \left[
    -\Delta \tau \hat{H}_0 \right] \ket*{\vec{R}^{(n-1)}}
  \nonumber \\
  & \times & \frac{\rho_T ( \vec{R}^{(n)}) }{\rho_T ( \vec{R}^{(n-1)})
  } F ( \vec{R}^{(0)},0 ) \ ,
  \label{itedmc}
\end{eqnarray}
where $\vec{R}^{(n)}$ 
are the position coordinates of the walker, and $s_k^{(n)}$ the spin
of particle $k$ of that walker, both at iteration $n$.
We can understand this expression in a simple way. The last two pieces
correspond to a standard GDB DMC
process~\cite{Guardiola} for the spinless part of the
Hamiltonian. On the other hand, the first part, incorporating the
spin-dependent terms, can be implemented through a
secondary branching process. This one must fulfill
that, after $j$ iterations, the weight carried by a given walker is
given by
\begin{equation}
w(j) = \sum_{\vec{S}^{(j)},...,\vec{S}^{(0)}} \prod_{n=1}^j \left(
  \prod_{k=1}^N \bra*{s^{(n)}_k} \hat{O}_k
  \ket*{s^{(n-1)}_k} \right) \ ,
\label{secondweight}  
\end{equation}
corresponding to the first term in Eq.~(\ref{itedmc}).  This is
fulfilled by performing the secondary branching at iteration $j$ using
the weight
\begin{equation}
 B(j) = \frac{ w(j) }{ w(j-1) }
\label{secondbranch}
\end{equation}
with the initial condition $w(0) = 1$. It can be shown
that $w(j)$ can be easily computed as
\begin{equation}
  w(j) = \prod_{k=1}^N \left( c_k^{+}(j) + c_k^{-}(j) \right)
  =\prod_{k=1}^N w_k(j)
  \ ,
  \label{weightc} 
\end{equation}
in terms of the spin weight factors
\begin{eqnarray}
 \mqty(c_k^{+}(j) \\ c_k^{-}(j) ) & = & \left[ \prod_{n=1}^j \mqty(
    \bra*{\uparrow}
    \hat{O}_k 
    \ket*{\uparrow} & \bra*{\uparrow}
    \hat{O}_k 
    \ket*{\downarrow} \\ \bra*{\downarrow}
    \hat{O}_k 
    \ket*{\uparrow} & \bra*{\downarrow}
    \hat{O}_k 
    \ket*{\downarrow} ) \right] 
    \mqty(1 \\ 1 ) \nonumber \\
 & = & \mqty( \bra*{\uparrow}
    \hat{O}_k 
    \ket*{\uparrow} & \bra*{\uparrow}
    \hat{O}_k 
    \ket*{\downarrow} \\ \bra*{\downarrow}
    \hat{O}_k 
    \ket*{\uparrow} & \bra*{\downarrow}
    \hat{O}_k 
    \ket*{\downarrow} ) \mqty(c_k^{+}(j-1) \\ c_k^{-}(j-1) )
    \label{defc}
\end{eqnarray}
where $\ket{\uparrow}$ and $\ket{\downarrow}$ stand for $\ket{s=1}$
and $\ket{s=-1}$, respectively. In this way, in the proposed method 
each walker carries the evolution of both $c^{+}$ and $c^{-}$ for
every particle, instead of explicit spin variables. 

Notice that the weights in Eq.~(\ref{weightc}) are the products of
one-body terms, due to the one-body nature of the spin-dependent
part of the interaction. Therefore, for each $k$, we can write
\begin{eqnarray}
  w(j) & = &
  c_k^+(j) \prod_{m\neq k} w_m(j)  + 
  c_k^-(j) \prod_{m\neq k} w_m(j) \nonumber \\ 
  & = &
  \eta_k^+(j) +
  \eta_k^-(j) \label{etadef} \ .
\end{eqnarray}
Equation~(\ref{secondweight}) can be expressed in terms of
$\eta_k^+(j)$ and $\eta_k^-(j)$, and we can 
rewrite Eq.~(\ref{itedmc}) as
\begin{eqnarray}
  F ( \vec{R}^{(j)},j\Delta \tau ) & = &
  \int \vec{dR}^{(j-1)}
  \cdot\cdot\cdot \vec{dR}^{(0)} ( \eta_k^+(j) +
  \eta_k^-(j) ) \nonumber \\
  & \times & \prod_{n=1}^j \bra*{\vec{R}^{(n)}} \exp \left[
    -\Delta \tau \hat{H}_0 \right] \ket*{\vec{R}^{(n-1)}}
  \nonumber \\
  & \times & \frac{\rho_T ( \vec{R}^{(n)}) }{\rho_T ( \vec{R}^{(n-1)})
  } F ( \vec{R}^{(0)},0 ) \ .
  \label{itedmc_new}
\end{eqnarray}
It can be shown that the marginal spin
integrated magnitude of the importance sampling function of
Eq.~(\ref{iterhois}) reads
\begin{eqnarray}
  \tilde F ( \vec{R}^{(j)},s_k=\pm 1,j\Delta \tau ) & = &
  \int \vec{dR}^{(j-1)}
  \cdot\cdot\cdot \vec{dR}^{(0)} \eta_k^{\pm}(j)
  \nonumber \\  
  & \times &
  \prod_{n=1}^j \bra*{\vec{R}^{(n)}} \exp \left[
    -\Delta \tau \hat{H}_0 \right] \ket*{\vec{R}^{(n-1)}}
  \nonumber \\
  & \times &  \frac{\rho_T ( \vec{R}^{(n)}) }{\rho_T ( \vec{R}^{(n-1)})
  } F( \vec{R}^{(0)},0 )
  \label{fmarg} \\
  & = &  \eval{ \sum_{\vec{S}_{N-l}} f(\vec{R}^{(j)},\vec{S}, j \Delta
    \tau) }_{s_k = \pm 1}
  \nonumber \ ,
\end{eqnarray}
where $\sum_{\vec S_{N-1}}$ in the second term means summing over the
spins of all particles but the $k$-th one.
This quantity is relevant in the energy
estimation, described in the next Section.

\subsection{\label{sec:localenergy}Energy estimation}

We show in this Section how to estimate the energy of a many-body
system under SOC interactions using the method introduced in the
previous Sections, although it can be easily extended to estimate any
other quantity.  The DMC energy estimator in the FPA at iteration $j$ is given by:
\begin{eqnarray}
 E_{\text{DMC}}(j) & = & \sum_{\vec{S},\vec{S}'} \int \vec{dR}^{(j)}
 \vec{dR'} \bra*{\vec{R}',\vec{S}'} \hat{H}^{\text{FP}}
 \ket*{\vec{R}^{(j)},\vec{S}} \nonumber \\
 & \times &
 \frac{\rho_T(\vec{R'}) }{ \rho_T(\vec{R}^{(j)})
 }f(\vec{R}^{(j)},\vec{S},j \Delta \tau) \ ,
 \label{edmc}
\end{eqnarray}
with $\hat{H}^{\text{FP}}$ defined in Eq.~(\ref{HFP}). The local
energy is, therefore,
\begin{eqnarray}
  E_L= \sum_{\vec{S}'} \int
 \vec{dR'} \bra*{\vec{R}',\vec{S}'} \hat{H}^{\text{FP}}
 \ket*{\vec{R}^{(j)},\vec{S}} \frac{ \rho_T(\vec{R'}) }{
   \rho_T(\vec{R}^{(j)}) } \ ,
\end{eqnarray}
which, as it can be seen, depends on $\vec R^{(j)}$ and $\vec{S}$,
so that $E_L=E_L(\vec{R}^{(j)},\vec{S})$. We can split it in two
parts
\begin{equation}
 E_L(\vec{R}^{(j)},\vec{S}) = E_{L,0}(\vec{R}^{(j)}) +
 E_{L,\text{S}}(\vec{R}^{(j)},\vec{S}) \ ,
\end{equation}
corresponding the the spin-independent and spin-dependent
contributions, respectively. The spin-independent part can be
expressed in the form
\begin{equation}
 E_{L,0}(\vec{R}^{(j)}) = \int \vec{dR'} \bra*{\vec{R}'}
 \hat{H}_0 \ket*{\vec{R}^{(j)}} \frac{ \rho_T(\vec{R'}) }{
   \rho_T(\vec{R}^{(j)}) } \ ,
  \label{elkinetic}
\end{equation}
while
\begin{eqnarray}
  E_{L,S} & = & \sum_{\vec{S}'} \int \vec{dR'}
  \bra*{\vec{R}',\vec{S}'} \hat{w}_{\text{Re}} + \hat{V}_{\Phi}
  \ket*{\vec{R}^{(j)},\vec{S}} 
  \frac{ \rho_T(\vec{R'}) }{ \rho_T(\vec{R}^{(j)}) }
  \nonumber \\
    & = & \sum_{l=1}^N \epsilon_{L,S,l}
  (\vec{R}^{(j)},s_l) \ ,
 \label{el1b}
\end{eqnarray}
with $\epsilon_{L,S,l}$ the one-body contribution to the spin-dependent local energy corresponding to particle $l$
(recall that $\hat w_{\rm Re}+\hat V_\Phi$ is a one-body operator).
With all these definitions,  Eq.~(\ref{edmc}) becomes
\begin{equation}
  E_{\text{DMC}}(j) = E_{\text{DMC},0}(j) + E_{\text{DMC},\text{S}}(j)
  \label{edmc2}
\end{equation}
The term $E_{\text{DMC},0}(j)$ contains all the spin-independent
contributions, and can be written as
\begin{eqnarray}
  E_{\text{DMC},0}(j) & = & \int \vec{dR}^{(j)} \text{ }
  E_{L,0}(\vec{R}^{(j)}) \sum_{\vec{S}}
  f(\vec{R}^{(j)},\vec{S},j \Delta \tau)
  \nonumber \\
  & = & \int \vec{dR}^{(j)} \text{ } E_{L,0}(\vec{R}^{(j)})
  F(\vec{R}^{(j)},j \Delta \tau)
  \label{edmc0}
\end{eqnarray}
with $F(\vec R,\tau)$ defined in Eq.~(\ref{intrho}). This part of the energy is evaluated as usual in DMC, i.e.
\begin{eqnarray}
  E_{\text{DMC},\text{0}}(j) & = & \frac{1}{ N_w }
  \sum_{i_w = 1}^{N_w}
  E^{(i_w)}_{L,\text{0}}(\vec{R}^{(j)})
  \ ,
 \label{edmc_spinless_est}
\end{eqnarray}
where $N_w$ is the total number of walkers in the simulation, and
$i_w$ specifies the walker index. In much the same
way
\begin{eqnarray}
  E_{\text{DMC},\text{S}}(j) & = & \sum_{l=1}^N \sum_{s_l=\pm 1}
  \int \vec{dR}^{(j)} \epsilon_{L,\text{S},l}(\vec{R}^{(j)},s_l)
  \nonumber \\  
  & \times & \tilde F ( \vec{R}^{(j)},s_l,j\Delta \tau ) \label{edmc1b}
\end{eqnarray}
with $\tilde F( \vec{R}^{(j)},s_l,j\Delta \tau )$ defined in
Eq.~(\ref{fmarg}). Therefore, we need to be able to sample $\tilde
F(\vec{R}^{(j)},s_l,j\Delta \tau )$ in order to evaluate
$E_{\text{DMC},\text{S}}(j)$. This can be done by estimating
$E_{\text{DMC},\text{S}}(j)$ as
\begin{eqnarray}
  E_{\text{DMC},\text{S}}(j) & = & \frac{1}{ N_w }
  \Big( \sum_{i_w = 1}^{N_w} \sum_{l=1}^N \frac{
    c_{l,i_w}^{ + }(j) }{ c_{l,i_w}^{ + }(j) + c_{l,i_w}^{ - }(j) }
  \epsilon^{(i_w)}_{L,\text{S},l}(\vec{R}^{(j)},+1)
  \nonumber \\
  & + & \frac{
    c_{l,i_w}^{ - }(j) }{ c_{l,i_w}^{ + }(j) + c_{l,i_w}^{ - }(j) }
  \epsilon^{(i_w)}_{L,\text{S},l}(\vec{R}^{(j)},-1) \Big)
  \nonumber \\
  & = & {1\over N_w} \sum_{i_w=1}^{N_w} \varepsilon^{(i_w)}_{L,\text{S}}(\vec{R}^{(j)})
  \ ,
 \label{edmc1best}
\end{eqnarray}
This expression ensures that each local energy contribution
$\epsilon^{(i_w)}_{L,\text{S},l}(\vec{R}^{(j)},\pm 1)$ is averaged
with an effective weight given by
\begin{align}
 &\frac{ c_{l}^{ \pm }(j) }{ c_{l}^{ + }(j) + c_{l}^{ - }(j)  } w(j) =
  \eta_l^{\pm}(j) \ ,
\end{align}
which is the one associated to $\tilde F ( \vec{R}^{(j)},s_l=\pm 1,j\Delta
\tau )$ in Eq.~(\ref{fmarg}). It is important to realize that
Eq.~(\ref{edmc1best}) can be used to estimate the expectation value of
any quantity that depends on the spin through one-body terms only,
replacing the $\epsilon_{L,S}(\vec R, \pm1)$ terms with the corresponding operators.

\subsection{\label{sec:algorithm}The SIDMC algorithm}

In this section we present a scheme of the Spin-integrated DMC
algorithm. In the present method, a walker is represented by the set
of quantities
\begin{equation}
  \vec{v} = \left( \vec{r}_1, \ldots, \vec{r}_N, c_1^+, c_1^-, \ldots,
  c_N^+, c_N^- \right) \ .
\end{equation}
Particle positions are initialized as usual in Monte Carlo
simulations, while spin weight factors $c_k^\pm$ must be initialized
to one in the first iteration
\begin{eqnarray}
 c_k^{ \pm } = 1 \text{ } \forall k \ .
\end{eqnarray}

The first step in each iteration of the algorithm is to perform a
standard GDB process using the
spinless part of the Hamiltonian $\hat{H}_0$ and
$\rho_T(\vec{R})$.
Next, one has to update the $c_k^{\pm}$ coefficients according to the expression 
\begin{equation}
 \mqty(c_k^{+}(j+1) \\ c_k^{-}(j+1) ) = \boldsymbol{O}^{(j+1)}_{k}
 \mqty(c_k^{+}(j) \\ c_k^{-}(j) ) \ ,
\end{equation}
which yields the new coefficients at iteration $j+1$ from the known
ones at iteration $j$. Notice that, in this expression,
$\boldsymbol{O}_k$ is the $2 \times 2$ matrix of Eq.~(\ref{defc}). 
Once with these coefficients, one can obtain $w(j+1)$
according to
\begin{equation}
 w(j+1) = \left[ \prod_{k=1}^N \left( c_k^{+}(j+1) + c_k^{-}(j+1)
   \right) \right] \ .
\label{branching_alg2}
\end{equation}
and from here, the secondary branching factor, 
\begin{equation}
B(j+1) = \frac{w(j+1)}{w(j)} \ .
\label{branching_alg}
\end{equation}
Notice this weight is different for each walker, so in fact
$B=B_{i_w}$ with $i_w$ the walker index. 

In practice, it may happen that, along the simulation, the absolute
value of the $c_k^{\pm}(j)$ coefficients keeps increasing
unboundedly. However, the ratio of $w$'s in this equation is always
finite. On the other hand, it is better to use a mixed-branching strategy
with the $B(j+1)$ terms, where walkers acquire a weight that is being
updated along each block of iterations. The accumulated weight $\mathcal{B}_{i_w}$
at the end of the block is equal to the product of the weights at each
iteration, for each walker.  Once the block is finished, these weights
are used to replicate the list of walkers.

In DMC simulations, the weight of the walkers is divided by a constant
(equal to $e^{E_T \Delta\tau}$ with $E_T$ the threshold energy and
$\Delta \tau$ the time step) when performing the replication
process~\cite{Guardiola}.
One has to perform an equivalent renormalization with the secondary
branching, while in this case the normalization constant can be
computed in two ways. One way is to use the average over the final
number of walkers of the accumulated $B$ of the previous block. 
Another way is to use the $B$ coefficients of the current block, 
accumulated over the previous iterations and averaged over the number of
walkers. The best strategy is determined by the SOC model at hand,
with the first choice being more suitable for the Raman interaction,
and the latter performing better with the Weyl and Rashba models. 

The energy at iteration $i$ inside a block is estimated as
\begin{eqnarray}
  E_{DMC}^{(i)} & = &
  {\sum_{i_w=1}^{N_w} E_{i_w}^{(i)} \mathcal{B}_{i_w}
    \over
  \sum_{i_w=1}^{N_w} \mathcal{B}_{i_w}}
  \\
  E_{i_w}^{(i)} & = & E^{(i_w)}_{L,\text{0}}(\vec{R}^{(j)}) + \varepsilon^{(i_w)}_{L,\text{S}}(\vec{R}^{(j)}) \ .
\end{eqnarray}
with $E^{(i_w)}_{L,\text{0}}(\vec{R}^{(j)})$ and
$\varepsilon^{(i_w)}_{L,\text{S}}(\vec{R}^{(j)})$ given in
Eqs.~(\ref{elkinetic}) and~(\ref{edmc1best}). In this expression the sum
is over the complete set of $N_w$ walkers, obtained after the standard
GDB process associated to the spinless part of the Hamiltonian. In
this way, the expression implicitly includes the weighting of the
standard branching. This equation represents the generalization of
Eqs.~(\ref{edmc_spinless_est}) and~(\ref{edmc1best}) for the 
mixed-branching case.

An important remark concerning the secondary branching is that
$B(j+1)$ in Eq.~(\ref{branching_alg}) is not positive definite. However,
the fraction of walkers which generate a change in sign is tiny, and
thus walkers that produce this effect can be safely discarded.
To quantify that, we monitor the quantity
\begin{equation}
\chi = \frac{N_e}{\langle N_w \rangle N_b} \ ,
\end{equation}
with $N_e$ and $N_b$ the number of eliminated walkers and the number
of iterations per block, and $\langle N_w\rangle$ the average number
of walkers of the block. Our numerical results show that $\chi$
depends slightly on the value of the parameters chosen for the
simulation, but it is always of the order of $10^{-3}$ or smaller.

\section{\label{sec:tmoves}Discrete spin T-moves DMC (DTDMC)}

In this section we adapt the continuous spin T-moves method of
Ref.~\cite{mitas} to a system of discrete spins under the SOC
interactions analyzed in this work. In the following, we assume the
two-body interaction is spin-dependent, with (possibly) different
contributions in each channel.  In this method the walkers carry explicit
spin variables together with the particle positions.

\subsection{\label{sec:tmovesA}Formalism}

In order to derive the alternative algorithm, one has to go back to
the beginning and work out the propagator in Eq.~(\ref{propagator_FP}),
which we split in a different way rearranging terms as follows
\begin{eqnarray}
  G_{\text{FP}} ( \vec{R},\vec{S} & \rightarrow & \vec{R}',\vec{S}') =   \int \vec{dR''} \bra*{\vec{R}',\vec{S}'} \exp \left[ -\Delta \tau
    \hat{w}_{\text{Re}} \right] \ket*{\vec{R}'',\vec{S}} \nonumber \\
 & \times &  \!\!\bra*{\vec{R}'',\vec{S}} \exp \left[ -\Delta \tau \hat{H}_{1}
    \right] \ket*{\vec{R},\vec{S}} + \order{\Delta \tau^2} \ ,
  \label{propagatorFP}
\end{eqnarray}
where
\begin{equation}
 \hat{H}_{1} = \sum_{k=1}^N \left[ \frac{P_k^2}{2M} +
   \hat{V}_{k}^{\text{1b}} + \abs*{ \vec{\nabla}_k
     \Phi_T(\vec{R},\vec{S}) }^2 \right] + \hat{V}^{\text{2b}} \ .
 \label{Hsl}
\end{equation}
We can introduce the importance sampling function inside this
expression and write
\begin{eqnarray}
 \frac{\rho_T(\vec{R}',\vec{S}')}{\rho_T(\vec{R},\vec{S})}
 G_{\text{FP}} ( \vec{R},\vec{S} \rightarrow \vec{R}',\vec{S}') & = & 
  \label{propagatorFP} \\
 \int \vec{dR''}
 \frac{\rho_T(\vec{R}',\vec{S}')}{\rho_T(\vec{R}'',\vec{S})}
 \bra*{\vec{R}',\vec{S}'} \exp \Big[ \!& - & \!\Delta \tau \hat{w}_{\text{Re}}
   \Big] \ket*{\vec{R}'',\vec{S}} \nonumber \\
 \times \frac{\rho_T(\vec{R}'',\vec{S})}{\rho_T(\vec{R},\vec{S})}
 \bra*{\vec{R}'',\vec{S}} \exp \Big[ \!&-&\!\Delta \tau \hat{H}_{1} \Big] \ket*{\vec{R},\vec{S}} + \order{\Delta \tau^2}
 \nonumber
\end{eqnarray}
To order $\order{\Delta \tau}$, the first term inside the integral becomes
\begin{eqnarray}
  \frac{\rho_T(\vec{R}',\vec{S}')}{\rho_T(\vec{R}'',\vec{S})}
  \bra*{\vec{R}',\vec{S}'} \exp \Big[ \!&-&\!\Delta \tau
    \hat{w}_{\text{Re}} \Big] \ket*{\vec{R}'',\vec{S}}
  \label{trans2} \\
  \simeq
  \delta(\vec{R}'-\vec{R}'') \delta(\vec{S}'-\vec{S}) 
  & - & \Delta \tau \bra*{\vec{R}',\vec{S}'} \hat{w}_{\text{Re}}
  \ket*{\vec{R}'',\vec{S}} \frac{\rho_T(\vec{R}',\vec{S}')}{\rho_T(\vec{R}'',\vec{S})}
  \nonumber 
\end{eqnarray}
However, for any kind of spin-orbit coupling the matrix element
$\bra*{\vec{R}',\vec{S}'} \hat{w}_{\text{Re}}
\ket*{\vec{R}'',\vec{S}}$ is not always negative, and thus Eq.~(\ref{trans2})
can not be interpreted as a probability distribution. In order to
bypass this limitation and in the spirit of Refs.~\cite{mitas, tmoves, ceperley},
we define an effective Hamiltonian that replaces the original one, and
that leads to a variational upper bound to the fixed phase energy of
the original Hamiltonian. We thus write
\begin{equation}
  \hat{H}^{\text{FP}}_{\text{eff}} = \hat{H}_{1} +
  \hat{w}_{\text{Re,A}}^{\text{eff}} +
  \hat{w}_{\text{Re,B}}^{\text{eff}} \ ,
  \label{effH_1}
\end{equation}
where the sum $\hat{w}_{\text{Re,A}}^{\text{eff}} +
\hat{w}_{\text{Re,B}}^{\text{eff}}$ is an approximation to the
original $\hat w_{\rm Re}$ of Eq.~(\ref{w_RE_guais}).  This
approximation is built such that the local energy of $\hat H_{\rm
  eff}^{\rm FP}$ and $\hat H^{\rm FP}$ are equal when they act on the
magnitude of the trail wave function. The matrix elements of these
terms are given by
\begin{eqnarray}
  \bra*{\vec{R},\vec{S}} \hat{w}_{\text{Re,A}}^{\text{eff}}
  \ket*{\vec{R},\vec{S}} & = & 0
  \label{effH_3} \\
  \bra*{\vec{R}',\vec{S}'} \hat{w}_{\text{Re,A}}^{\text{eff}}
  \ket*{\vec{R},\vec{S}}
  & = &
  \begin{cases}
    \bra*{\vec{R}',\vec{S}'} \hat{w}_{\text{Re}}
    \ket*{\vec{R},\vec{S}} &
    \text{ if } T < 0   \\
     0 & \text{ if } T > 0 
  \end{cases} \nonumber 
\end{eqnarray}
with the transition coefficients
\begin{equation}
  T = 
  \bra*{\vec{R}',\vec{S}'}
  \hat{w}_{\text{Re}} \ket*{\vec{R},\vec{S}}
  \frac{\rho_T(\vec{R}',\vec{S}')}{\rho_T(\vec{R},\vec{S})} \ ,
  \label{effH_2}
\end{equation}
while
\begin{eqnarray}
 \bra*{\vec{R},\vec{S}} \hat{w}_{\text{Re,B}}^{\text{eff}}
 \ket*{\vec{R},\vec{S}} & = & \sum_{\vec{s}} \int \vec{dX}
 \bra*{\vec{R},\vec{S}} \hat{w}_{\text{Re}}
 \ket*{\vec{X},\vec{s}}
 \frac{\rho_T(\vec{X},\vec{s})}{\rho_T(\vec{R},\vec{S})}
 \nonumber \\
 \bra*{\vec{R}',\vec{S}'} \hat{w}_{\text{Re,B}}^{\text{eff}}
 \ket*{\vec{R},\vec{S}} & = & 0
 \label{effH_6}
\end{eqnarray}
where in the last expression, the summation and the integration are
restricted to those values that satisfy the condition $T>0$.
Using these definitions we avoid non-local matrix elements producing negative
transition probabilities.
Notice also that the effective Hamiltonian depends on the magnitude of
the trial wave function, which means that the the energy obtained
depends on its choice.  We showcase this effect in
Sec. \ref{sec:results}. The fixed-phase propagator
for the effective Hamiltonian, with importance sampling, is thus:
\begin{eqnarray}
  \frac{\rho_T(\vec{R}',\vec{S}')}{\rho_T(\vec{R},\vec{S})}
  G_{\text{FP}}^{\text{eff}} ( \vec{R},\vec{S} \rightarrow
  \vec{R}',\vec{S}') & = & \nonumber \\ 
  \times \int \vec{dR''}
  \frac{\rho_T(\vec{R}',\vec{S}')}{\rho_T(\vec{R}'',\vec{S})}
  \bra*{\vec{R}',\vec{S}'} \exp \Big[ \!&-&\!\Delta \tau
    \hat{w}_{\text{Re,A}}^{\text{eff}} \Big]
  \ket*{\vec{R}'',\vec{S}} \nonumber \\
  \times \frac{\rho_T(\vec{R}'',\vec{S})}{\rho_T(\vec{R},\vec{S})}
  \bra*{\vec{R}'',\vec{S}} \exp \Big[\!&-&\!\Delta \tau (\hat{H}_{1} +
    \hat{w}_{\text{Re,B}}^{\text{eff}}) \Big] \ket*{\vec{R},\vec{S}}
  \nonumber \\
  & & + \order{\Delta \tau^2} \ .
 \label{propagatorFP_effH}
\end{eqnarray}
Since this propagator is positive-definite, we can now interpret it as a probability distribution. Therefore, one can sample from it. This can be
implemented performing initially a GDB of the
$\exp[-\Delta\tau(\hat H_1 + \hat w_{\rm Re,B}^{\rm eff})]$ part, with
a branching factor that, according to Ref.~\cite{tmoves}, reads
\begin{equation}
 B(\vec{R},\vec{R}'', \vec{S}) = \exp \left[ -\frac{\Delta
     \tau}{2} \left[ E_L(\vec{R},\vec{S}) + E_L(\vec{R}'',\vec{S} )
     \right] \right] \ ,
  \label{branchtot_effH}
\end{equation}
with
\begin{equation}
  E_L(\vec{R},\vec{S}) = \sum_{\vec{S}'} \int \vec{dR'}
  \bra*{\vec{R}',\vec{S}'} \hat{H}_{\text{eff}}^{\text{FP}}
  \ket*{\vec{R},\vec{S}} \frac{ \rho_T(\vec{R'},\vec{S'}) }{
    \rho_T(\vec{R},\vec{S}) } \ ,
  \label{localenergy_tmoves_eq}
\end{equation}
which generates the displacement $\vec{R}
\rightarrow \vec{R}''$. In a second step, one performs 
a transition $(\vec{R}'',\vec{S})
\rightarrow (\vec{R}',\vec{S}')$ given by the probability
\begin{equation}
  p(\vec{R}'', \vec{S} \rightarrow \vec{R}' \vec{S}') =
  \frac{ P(\vec{R}'', \vec{S} \rightarrow \vec{R}' \vec{S}') }{
    \sum_{\vec{S}'} \int \vec{dR'} P(\vec{R}'', \vec{S} \rightarrow
    \vec{R}' \vec{S}') } \ ,
 \label{trans2_final_effH_A}
\end{equation}
where
\begin{eqnarray}
 P(\vec{R}'', \vec{S} & \rightarrow & \vec{R}' \vec{S}') =
 \delta(\vec{R}'-\vec{R}'') \delta(\vec{S}'-\vec{S}) \nonumber \\
 &-& \!\Delta \tau \bra*{\vec{R}',\vec{S}'}
 \hat{w}_{\text{Re,A}}^{\text{eff}} \ket*{\vec{R}'',\vec{S}}
 \frac{\rho_T(\vec{R}',\vec{S}')}{\rho_T(\vec{R}'',\vec{S})} \ .
  \label{trans2_effH}
\end{eqnarray}
Despite the sum in Eq.~(\ref{trans2_final_effH_A}) involves the $2^N$
spin configurations, which sounds prohibitive for large $N$, it must be
kept in mind that only one-body operators are involved and the
expression is greatly simplified.

\subsection{\label{sec:tmovesB}Application to synthetic SOC in ultracold gases}

In this section we particularize the results of the previous formalism
to the Weyl SOC interaction. The procedure is analogous for the Rashba
and Raman potentials. We start evaluating the matrix elements of
$\hat{w}_{\text{Re}}$, which are given by
\begin{eqnarray}
  \bra*{\vec{R}',\vec{S}'} \hat{w}_{\text{Re}} \ket*{\vec{R},\vec{S}}
   & = &{\lambda \hbar \over M} \sum_{k=1}^N \left[ \prod_{l \neq k} \delta(\vec{r}'_l -
    \vec{r}_l) \delta(s'_l - s_l) \right] \nonumber \\
   \times \Bigg[
    \delta(y_k'&-&y_k) \delta(z_k'-z_k)
          {d \over dx_k'} \delta(x_k'-x_k) \nonumber \\
          \times \bra*{s_k'} \hat{\sigma}_{x,k} \ket{s_k}
          \sin \Big[ \!&-&\!\phi_k(x_k',y_k,z_k,s_k') +
            \phi_k(\vec{r}_k,s_k) \Big]
          \nonumber \\
    + \delta(x_k'&-&x_k) \delta(z_k'-z_k)
          {d \over dy_k'} \delta(y_k'-y_k) \nonumber \\
          \times \bra*{s_k'} -i\sigma_{y,k} \ket{s_k}
          \cos \Big[ \!&-&\!\phi_k(x_k,y_k',z_k,s_k') +
            \phi_k(\vec{r}_k,s_k) \Big]
          \nonumber \\          
    + \delta(x_k'&-&x_k) \delta(y_k'-y_k)
    {d \over dz_k'} \delta(z_k'-z_k)
    \label{wRe_churro} \\
          \times \bra*{s_k'} \sigma_{z,k} \ket{s_k}
          \sin \Big[ \!&-&\!\phi_k(x_k,y_k,z_k',s_k') +
            \phi_k(\vec{r}_k,s_k) \Big]
          \nonumber
\end{eqnarray}
with $\phi_k$ the single-particle phase of Eq.(~\ref{trialphase}).  In
this expression we have omitted the last term of Eq.~(\ref{Wweyl}) as
it is a constant contribution that represents a shift of the total
energy only.  In order to construct the effective Hamiltonian, we must
evaluate the matrix elements of $\hat{w}_{\text{Re}}$ to check their
sign. However, given any set of coordinates $\vec{r}_k$, $\vec{r}'_k$,
terms of the form $\dv{}{x_k'}\left( \delta(x_k' - x_k) \right)$ are
in general problematic. In order to preserve the upper bound property
of the effective Hamiltonian, we adopt the (apparently rude)
prescription 
\begin{equation}
 \dv{}{\xi_k'}\left( \delta(\xi_k' - \xi_k) \right) \sim
 \frac{1}{2\epsilon} \left[ \delta(\xi_k' + \epsilon - \xi_k) -
   \delta(\xi_k' - \epsilon - \xi_k) \right]
 \label{deltas_chachis}
\end{equation}
with $\epsilon$ a small parameter. This is equivalent to replacing the
momentum operator with
\begin{equation}
 \hat{p} \sim \frac{\hbar}{2 i \epsilon} \left[ \exp \left( i
   \frac{\hat{p}}{\hbar} \epsilon \right) - \exp \left( -i
   \frac{\hat{p}}{\hbar} \epsilon \right) \right]
 \label{subs_op}
\end{equation}
while both expressions coincide to order $\epsilon$. Notice that, in this form, the
resulting operator is still hermitian, and for $\epsilon\to 0$, the
energy is preserved. With this substitution, $\hat w_{\rm Re}$ is
replaced by a new operator $\hat{w}_{\text{Re},\epsilon}$,
whose matrix elements are the same as in Eq.~(\ref{wRe_churro}) with
the derivatives of the deltas replaced as in
Eq.~(\ref{deltas_chachis}). 
We can now construct the effective Hamiltonian using the definitions
in Eqs.~(\ref{effH_1} - \ref{effH_6}), with $\hat w_{\rm Re,\epsilon}$
replacing $\hat w_{\rm Re}$, which give raise to the
effective Hamiltonian contributions
$\hat{w}_{\text{Re},\epsilon,\text{A}}^{\text{eff}}$ and
$\hat{w}_{\text{Re},\epsilon,\text{B}}^{\text{eff}}$.

Notice that, by introducing the prescription in
Eqs.~{\ref{deltas_chachis}) and~(\ref{subs_op}), the SOC part of the propagator
  becomes exact up to order $\order{\frac{N \Delta \tau}{2\epsilon}}$.
  This implies that the value of $\epsilon$ must be chosen so that
\begin{eqnarray}
 1 & \gg & \abs{ \Delta \tau \sum_{\vec{S}'} \int \vec{dR'}
   \bra*{\vec{R}',\vec{S}'}
   \hat{w}_{\text{Re},\epsilon,\text{A}}^{\text{eff}}
   \ket*{\vec{R}'',\vec{S}}
   \frac{\rho_T(\vec{R}',\vec{S}')}{\rho_T(\vec{R}'',\vec{S})} }
 \nonumber 
 \\
 1 & \gg & \abs{ \Delta \tau \sum_{\vec{S}'} \int \vec{dR'}
   \bra*{\vec{R}',\vec{S}'}
   \hat{w}_{\text{Re},\epsilon,\text{B}}^{\text{eff}}
   \ket*{\vec{R}'',\vec{S}}
   \frac{\rho_T(\vec{R}',\vec{S}')}{\rho_T(\vec{R}'',\vec{S})} }
 \nonumber \\
 & = & \abs{ \Delta \tau \bra*{\vec{R}'',\vec{S}}
   \hat{w}_{\text{Re},\epsilon,\text{B}}^{\text{eff}}
   \ket*{\vec{R}'',\vec{S}} } \ ,
 \label{condition_eps_exp_2}
\end{eqnarray}
though in our simulations we have seen that these conditions can be
somewhat relaxed.  In any case, the precise value of $\epsilon$
chosen for the simulations should not affect the energy contribution
from the SOC part of the Hamiltonian,
\begin{eqnarray}
\sum_{\vec{S}'} \int & \vec{dR'} &  \bra*{\vec{R}',\vec{S}'}
\hat{w}_{\text{Re},\epsilon, \text{A}}^{\text{eff}} +
\hat{w}_{\text{Re},\epsilon, \text{B}}^{\text{eff}}
\ket*{\vec{R},\vec{S}} \frac{ \rho_T(\vec{R'},\vec{S'}) }{
  \rho_T(\vec{R},\vec{S}) }
\simeq \nonumber \\
\sum_{\vec{S}'} \int & \vec{dR'} & \bra*{\vec{R}',\vec{S}'}
\hat{w}_{\text{Re}} \ket*{\vec{R},\vec{S}} \frac{
  \rho_T(\vec{R'},\vec{S'}) }{ \rho_T(\vec{R},\vec{S}) } \ .
 \label{condition_eps_el}
\end{eqnarray}

\subsection{\label{sec:algorithm_tmoves}The DTDMC algorithm}

We discuss in this Section a scheme of the DTDMC algorithm to better understand its practical implementation. A walker at iteration $j$ is described by

\begin{align}
 &\vec{v}(j) = \left( \vec{r}_1^{(j)}, s_1^{(j)}, \ldots, \vec{r}_N^{(j)}, s_N^{(j)} \right)
\end{align}

with $s_k = \pm 1$ the z-component of the spin of particle $k$ and subindexes and superindexes standing particles and iterations, respectively. The initial condition for the position and spin coordinates is generally  obtained through the sampling of the trial wave function using the Metropolis algorithm. 

The first step to be implemented at each iteration is a GDB process with the branching factor given by Eq.~(\ref{branchtot_effH}), which produces a spatial translation $\vec{R}^{(j)} \rightarrow \vec{R}^{(j)}_A$. After this, we need to sample the part of the propagator which depends on the effective potential $\hat{w}_{\text{Re},\epsilon,\text{A}}$. In this second step, a transition $ (\vec{R}^{(j)}_A, \vec{S}^{(j)} ) \rightarrow (\vec{R}^{(j+1)}, \vec{S}^{(j+1)} ) $ is performed given by the probability

\begin{align}
 &p(\vec{R}, \vec{S} \rightarrow \vec{R}' \vec{S}') = \frac{ P(\vec{R}, \vec{S} \rightarrow \vec{R}' \vec{S}') }{ \sum_{\vec{S}'} \int \vec{dR'} P(\vec{R}, \vec{S} \rightarrow \vec{R}' \vec{S}') }
 \label{trans2_final_effH}
 \\
 &P(\vec{R}, \vec{S} \rightarrow \vec{R}' \vec{S}') = \delta(\vec{R}'-\vec{R}) \delta(\vec{S}'-\vec{S}) \nonumber \\
 &- \Delta \tau \bra*{\vec{R}',\vec{S}'} \hat{w}_{\text{Re},\epsilon,\text{A}}^{\text{eff}} \ket*{\vec{R},\vec{S}} \frac{\rho_T(\vec{R}',\vec{S}')}{\rho_T(\vec{R},\vec{S})}
  \label{trans2_effH}
\end{align}

where we can identify $\vec{R} = \vec{R}^{(j)}_A$, $\vec{S} = \vec{S}^{(j)}$, $\vec{R}' = \vec{R}^{(j+1)}$ and $\vec{S}' = \vec{S}^{(j+1)}$. As an example, we explicitly report how this evolution is carried out for the Weyl SOC case. A possible transition probability is:
{\small
\begin{align}
 &P(\vec{R}, \vec{S} \rightarrow \vec{R}', \vec{S}') = \delta(\vec{R}'-\vec{R}) \delta(\vec{S}'-\vec{S}) \nonumber \\
 &- \Delta \tau \left\{ \sum_{k=1}^N \left[ \prod_{l \neq k} \delta(\vec{r}'_l - \vec{r}_l) \delta(s'_l - s_l) \right] \right. \nonumber \\
 &\left. \cross \frac{\lambda \hbar}{M} \left[ \delta(y_k' - y_k)\delta(z_k' - z_k) \frac{1}{2 \epsilon} \delta(x_k' + \epsilon - x_k) \right. \right. \nonumber \\
 &\left. \left. \cross \bra*{s_k'} \hat{\sigma}_{x,k} \ket{s_k} \sin \left[ -\phi_k(x_k',y_k,z_k,s_k') + \phi_k(\vec{r}_k,s_k) \right] \right. \right. \nonumber \\
 &\left. \left. - \delta(x_k' - x_k)\delta(z_k' - z_k) \frac{1}{2 \epsilon} \delta(y_k' - \epsilon - y_k) \right. \right. \nonumber \\
 &\left. \left. \cross \bra*{s_k'} -i\hat{\sigma}_{y,k} \ket{s_k} \cos \left[ -\phi_k(x_k,y_k',z_k,s_k') + \phi_k(\vec{r}_k,s_k) \right] \right. \right. \nonumber \\
 &\left. + \delta(x_k' - x_k)\delta(y_k' - y_k) \frac{1}{2 \epsilon} \left( \delta(z_k' + \epsilon - z_k) - \delta(z_k' - \epsilon - z_k) \right) \right. \nonumber \\
 &\left. \left. \cross \bra*{s_k'} \hat{\sigma}_{z,k} \ket{s_k} \sin \left[ -\phi_k(x_k,y_k,z_k',s_k') + \phi_k(\vec{r}_k,s_k) \right] \right] \right\} \nonumber \\
 &\cross \frac{\rho_T(\vec{R}',\vec{S}')}{\rho_T(\vec{R},\vec{S})}
\end{align}
 }
 Notice that the terms appearing in $P(\vec{R}, \vec{S} \rightarrow \vec{R}', \vec{S}')$ are different for each walker and each iteration. In general, one has to keep here only those terms of Eq.~(\ref{wRe_churro})
 (after the substitution of Eqs.~(\ref{deltas_chachis}) and~(\ref{subs_op})) that are strictly negative.
 This total transition probability is the sum of different transition probabilities $P_{t,k}^{(m)}$, so it can be written as
  {\small
 \begin{align}
 &= P_{t,k}^{(0)}(\vec{R}, \vec{S} \rightarrow \vec{R} \vec{S}) \delta(\vec{R}'-\vec{R}) \delta(\vec{S}'-\vec{S}) \nonumber \\
 &+ \sum_{k=1}^N \left[ \prod_{l \neq k} \delta(\vec{r}'_l - \vec{r}_l) \delta(s'_l - s_l) \right] \left\{ \delta(y_k' - y_k)\delta(z_k' - z_k) \right. \nonumber \\
 &\left. \cross \delta(x_k' + \epsilon - x_k) P_{t,k}^{(1)}(x_k, s_k \rightarrow x_k - \epsilon, -s_k) \right. \nonumber \\
 & \left. + \delta(x_k' - x_k)\delta(z_k' - z_k) \delta(y_k' - \epsilon - y_k) P_{t,k}^{(2)}(y_k, s_k \rightarrow y_k + \epsilon, -s_k) \right. \nonumber \\
 &\left. + \delta(x_k' - x_k)\delta(y_k' - y_k) \left( \delta(z_k' + \epsilon - z_k) P_{t,k}^{(3)}(z_k, s_k \rightarrow z_k - \epsilon, s_k) \right. \right. \nonumber \\
 &\left. \left. + \delta(z_k' - \epsilon - z_k) P_{t,k}^{(4)}(z_k, s_k \rightarrow z_k + \epsilon, s_k) \right) \right\}
  \label{trans_example}
\end{align}
}%
The probabilities $P_{t,k}^{(m)}$ depend on the coordinates of all particles but we only make explicit the dependence on the coordinates that change under each transition for the sake of simplicity. Notice that in this example there are $4N + 1$ possible transitions. We define the cumulative distribution vector as
\begin{align}
 &v_c(i_c) = \frac{ \sum_{i=1}^{i_c} v_2(i) }{ \sum_{i=1}^{4N+1} v_2(i) } \text{ , }i_c = 1,...,4N+1 & v_c(0) = 0
\end{align}
with
\begin{align}
 &v_2 = (1, P_{t,1}^{(1)},P_{t,1}^{(2)},P_{t,1}^{(3)},P_{t,1}^{(4)},...,P_{t,N}^{(1)},P_{t,N}^{(2)},P_{t,N}^{(3)},P_{t,N}^{(4)}) \ .
\end{align}
Notice that $v_c(i_c) \in (0,1] \text{ } \forall i_c$. To sample this discrete probability distribution function we follow the standard procedure: we generate a random number $\xi \in [0,1]$ and select the component of $v_c(i_{\text{trans}})$ that verifies 
\begin{align}
 &v_c(i_{\text{trans}}-1) < \xi \nonumber \\
  &v_c(i_{\text{trans}}) > \xi
\end{align}
Finally, we perform the transition associated to the quantity $v_2(i_{\text{trans}}) = v_c(i_{\text{trans}}) - v_c(i_{\text{trans}}-1)$, i.e., if $v_2(i_{\text{trans}}) = P_{t,k}^{(2)}$, the spin of particle $k$ flips and its coordinates are modified according to $x_k' = x_k$, $y_k' = y_k + \epsilon$, $z_k' = z_k$, while the rest of the system is left unchanged.

\section{\label{sec:results}Results}

We report in this Section results for the energy in different systems
for both the SIDMC and DTDMC methods. In Sec. \ref{sec:fewbody}, we
show the energy of a few one-body and two-body problems, while in
Sec. \ref{sec:manybody}, we report results for the energy of a few
many-body systems, both in the mean-field regime and out of it.
As a check of validity of the two DMC algorithms
for SOC systems, we compare the DMC estimations with energies obtained
from the imaginary-time evolution of the Schr\"{o}dinger equation (one
and two-body cases) and the Gross-Pitaevskii equation (many-body in
the dilute regime). We also comment on the technical issues mentioned in
Secs. \ref{sec:algorithm} and \ref{sec:algorithm_tmoves}, mainly the
elimination of walkers in SIDMC and the influence of the parameter
$\epsilon$ in DTDMC, as well as the dependence of the energy
estimation on the time step. In all cases, the parameters of the
Hamiltonian and the trial wave function are reported in reduced units
(see Sec. \ref{sec:units}).

\subsection{\label{sec:fewbody}One and two-body problems}

In this Section, we report DMC results for the energy corresponding to
four different physical situations: a three-dimensional (3D) one-body
system with Weyl SOC, a 3D one-body system with Raman SOC, and two
interacting two-dimensional (2D) two-body systems with Rashba SOC, one
featuring a spin-independent two-body interaction and another with a
spin-dependent one. All systems are harmonically confined.  We
summarize our results in Table~\ref{tab:table_fewbody}, which includes
the DMC energies obtained with both algorithms together with the
imaginary time evolution (ITE) estimates, both for the fixed-phase
Hamiltonian (Eq.~(\ref{HFP})) and the fixed-phase, effective Hamiltonian
(Eq.~(\ref{effH_1})). All SIDMC energies are obtained by performing
several simulations, changing the parameter $\Delta \tau$, and then
extrapolating the energy to the limit $\Delta \tau \rightarrow 0$. In
the Weyl and Rashba cases with DTDMC, one must carry out several
calculations changing $\Delta \tau$ and $\epsilon$ and then
extrapolate to the limits $\Delta \tau \rightarrow 0$, $\epsilon
\rightarrow 0$, and $\frac{\Delta \tau}{\epsilon} \rightarrow 0$. We
discuss below how to perform the triple limit involving $\Delta \tau$,
$\epsilon$, and $\frac{\Delta \tau}{\epsilon}$. This setup is not
necessary in the Raman calculations since the SOC part of the
propagator scales as $\order{N \Delta \tau}$ if $\epsilon$ is
sufficiently small.

\begin{figure}[t]
\centering
\includegraphics[width=0.85\linewidth]{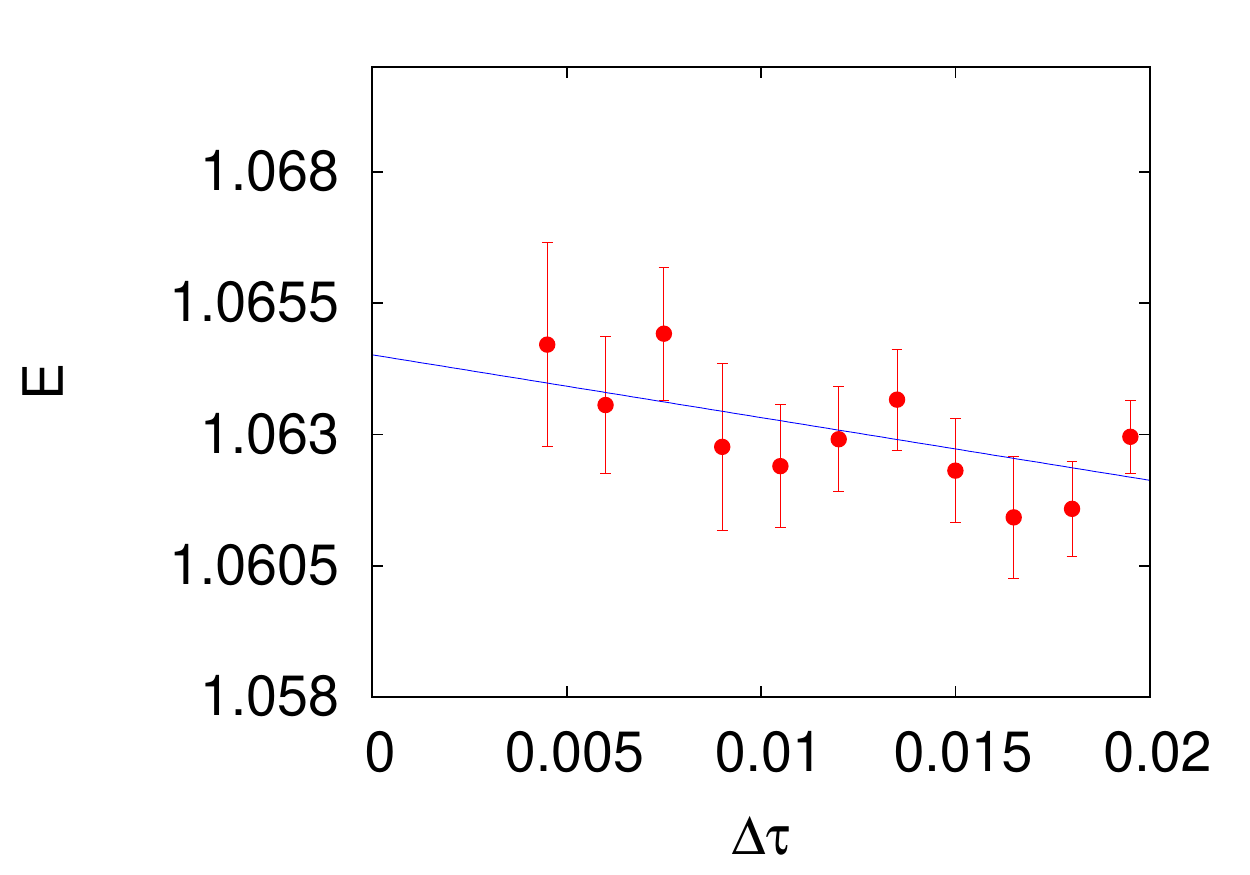}
\caption{Dependence of the DMC energy on the imaginary-time step using
  the SIDMC method for a two-body system with Rashba SOC and harmonic
  confinement.}
\label{fewbody_dt_jsanchez_rashba}
\end{figure}  

The trial wave function for each Hamiltonian is important because it
fixes the phase and, in all cases, reduces the variance via importance
sampling. In the problem of Raman SOC and DTDMC the trial wave
function that we have used is

\begin{align}
 &\Psi_T(\vec{r},s) = \rho_T(\vec{r},s) \exp \left[ i \phi_T (\vec{r},s) \right]
 \\
 &\rho_T(\vec{r},s=+1) = \left[ C_1^2 \sin^2 \mu + C_2^2 \cos^2 \mu \right. \nonumber \\
 &\left. + 2 \sin \mu \cos \mu C_1 C_2 \cos (2 k x) \right]^{1/2} \exp \left[ -\frac{\omega}{2} \left( x^2 + y^2 + z^2 \right) \right]
 \label{raman_mod_up}
 \\
 &\rho_T(\vec{r},s=-1) = \left[ C_2^2 \sin^2 \mu + C_1^2 \cos^2 \mu \right. \nonumber \\
 &\left. + 2 \sin \mu \cos \mu C_1 C_2 \cos (2 k x) \right]^{1/2} \exp \left[ -\frac{\omega}{2} \left( x^2 + y^2 + z^2 \right) \right]
 \label{raman_mod_down}
 \\
 &\phi_T(\vec{r},s=+1) = \atan \left[ \frac{ ( C_1 \sin \mu - C_2 \cos \mu ) \sin(k x) }{ ( C_1 \sin \mu + C_2 \cos \mu ) \cos(k x) } \right]
 \label{raman_phase_up}
 \\
 &\phi_T(\vec{r},s=-1) = \atan \left[ \frac{ ( C_1 \cos \mu - C_2 \sin \mu ) \sin(k x) }{ ( C_1 \cos \mu + C_2 \sin \mu ) \cos(k x) } \right]
 \label{raman_phase_down}
\end{align}
with $\mu = \frac{1}{2} \acos \left( \frac{k}{ \eta_{\text{Rm}} }
\right)$, $k$ the reduced momentum and $\omega$ the
harmonic oscillator strength. In these expressions, $\{k, C_1,C_2\}$ are taken as variational parameters. The SOC term of the trial wave function
is of the same form as the one used in Ref.~\cite{stringari}. Since
the magnitude of the trial wave function must be independent of the spin in SIDMC, we have used
\begin{align}
 &\rho_T(\vec{r}) = \left[ C_1^2 + C_2^2 + 2 B_c C_1 C_2 \cos (2 k x) \right]^{1/2} \nonumber \\
 & \cross \exp \left[ -\frac{\omega}{2} \left( x^2 + y^2 + z^2 \right) \right]
 \label{raman_mod}
\end{align}
with $B_c$ another variational parameter.

Concerning the Weyl model, the adopted trial wave function for DTDMC is 
\begin{align}
 &\rho_T(\vec{r},s=+1) = \exp \left[ -\frac{\omega}{2} \left( x^2 + y^2 + z^2 \right) \right]
 \label{weyl_mod_up}
 \\
 &\rho_T(\vec{r},s=-1) = \frac{ \left( 1 + \cos \theta_k \right) }{ \sin \theta_k } \exp \left[ -\frac{\omega}{2} \left( x^2 + y^2 + z^2 \right) \right]
 \label{weyl_mod_down}
 \\
 &\phi_T(\vec{r},s=+1) = \vec{k} \vec{r}
 \label{weyl_phase_up}
 \\
 &\phi_T(\vec{r},s=-1) = \vec{k} \vec{r} + \pi + \phi_k
 \label{weyl_phase_down}
\end{align}
where $\theta_k$ and $\phi_k$ are the polar and azimuthal angles of
the momentum vector $\vec{k}$, respectively. The adopted magnitude of the trial wave function for
the SIDMC case is
\begin{align}
 &\rho_T(\vec{r}) = \exp \left[ -\frac{\omega}{2} \left( x^2 + y^2 + z^2 \right) \right]
 \label{weyl_mod}
\end{align}
Finally, the trial wave function used in the DTDMC two-body Rashba simulations is 
\begin{align}
 &\Psi_T(\vec{R},\vec{S}) = \left[ \prod_{j=1}^2 \rho_{T,\text{1b}} (\vec{r}_j,s_j) \right] \rho_{T,\text{2b}} (\vec{r}_1,\vec{r}_2) \nonumber \\
 & \cross \exp \left[ i \sum_{j=1}^2 \phi_T (\vec{r}_j,s_j) \right]
 \\
 &\rho_{T,\text{1b}}(\vec{r},s=+1) = \exp \left[ -\frac{\omega}{2} \left( x^2 + y^2 \right) \right]
 \\
 &\rho_{T,\text{1b}}(\vec{r},s=-1) = \exp \left[ -\frac{\omega}{2} \left( x^2 + y^2 \right) \right]
 \\
 &\phi_T(\vec{r},s=+1) = \vec{k} \vec{r} - \phi_k - \frac{\pi}{2}
 \\
 &\phi_T(\vec{r},s=-1) = \vec{k} \vec{r}
\end{align}
with $\phi_k$ the angle of the momentum vector in polar coordinates. In this expression, $\rho_{T,\text{2b}}(\vec{r}_1,\vec{r}_2)$ is the exact
solution of the two-body interacting problem at low momentum
($k_{\text{2b}} \sim 10^{-2}$) (without SOC) corresponding to the
soft-sphere potential of Eq.~(\ref{V2b}), with parameters {\small
\begin{align}
 &\overline{V}_0 = \frac{ V_0(1,1) + V_0(1,-1)+ V_0(-1,1)+ V_0(-1,-1) }{4}
 \\
 &\overline{R}_0 = \frac{ R_0(1,1) + R_0(1,-1)+ R_0(-1,1)+ R_0(-1,-1) }{4} \ .
\end{align}
}
This choice makes the two-body trial wave function spin-independent for simplicity. We use the same choice
for the SIDMC simulations.

The time step is $\Delta \tau \sim \order{10^{-3}}$ in DTDMC
simulations while it is $\Delta \tau \sim \order{10^{-2}}$ in the
SIDMC ones.  The average number of walkers is kept stable along the
simulations, and it is fixed to a value between 2000 and 3000,
depending on the case.
The parameter $\epsilon$ of DTDMC is fixed as $\epsilon =
100 \Delta \tau$ in the Raman calculation and as $\epsilon = 200
\Delta \tau$ in the Rashba and Weyl cases. In the Weyl SIDMC
calculations, the secondary branching weights $w(j)$ are accumulated along
blocks of $N_b=10$ iterations.
The ratio of eliminated walkers is $\chi < 0.001$. In the Rashba
cases, we have $N_b = 50$ and $\chi < 0.002$. Finally,
for the Raman problem we have $N_b = 10$ and $\chi = 0$
(see Sec. \ref{sec:algorithm}).

The parameters used in the Raman simulations are $\eta_{\text{Rm}} =
1$, $\omega = 0.4$, $\Omega = 0.5$, $k = 0.7$, $C_1 = 0.6$, $C_2 =
0.8$, and $B_c = 0.5$. For the Weyl simulations we considered
$\eta_{\text{We}} = 1$, $\omega = 0.4$, $k = 0.5$, $\theta_k =
\frac{\pi}{4}$, and $\phi_k = 0.3$. Finally, the parameters for the
two-body Rashba simulations in the two-body spin-independent case are
$V_0 = 1.5$, $R_0 = 3.5$, $k = 0.5$, $\phi_k =0.1$, and $\omega =
0.4$. The two-body spin-dependent Rashba case shares the same values,
except for $V_0(+1,+1) = V_0(-1,-1) = 2.5$ and $V_0(+1,-1) =
V_0(-1,+1) = 1.5$.

\begin{figure}[b]
\centering
\includegraphics[width=0.85\textwidth]{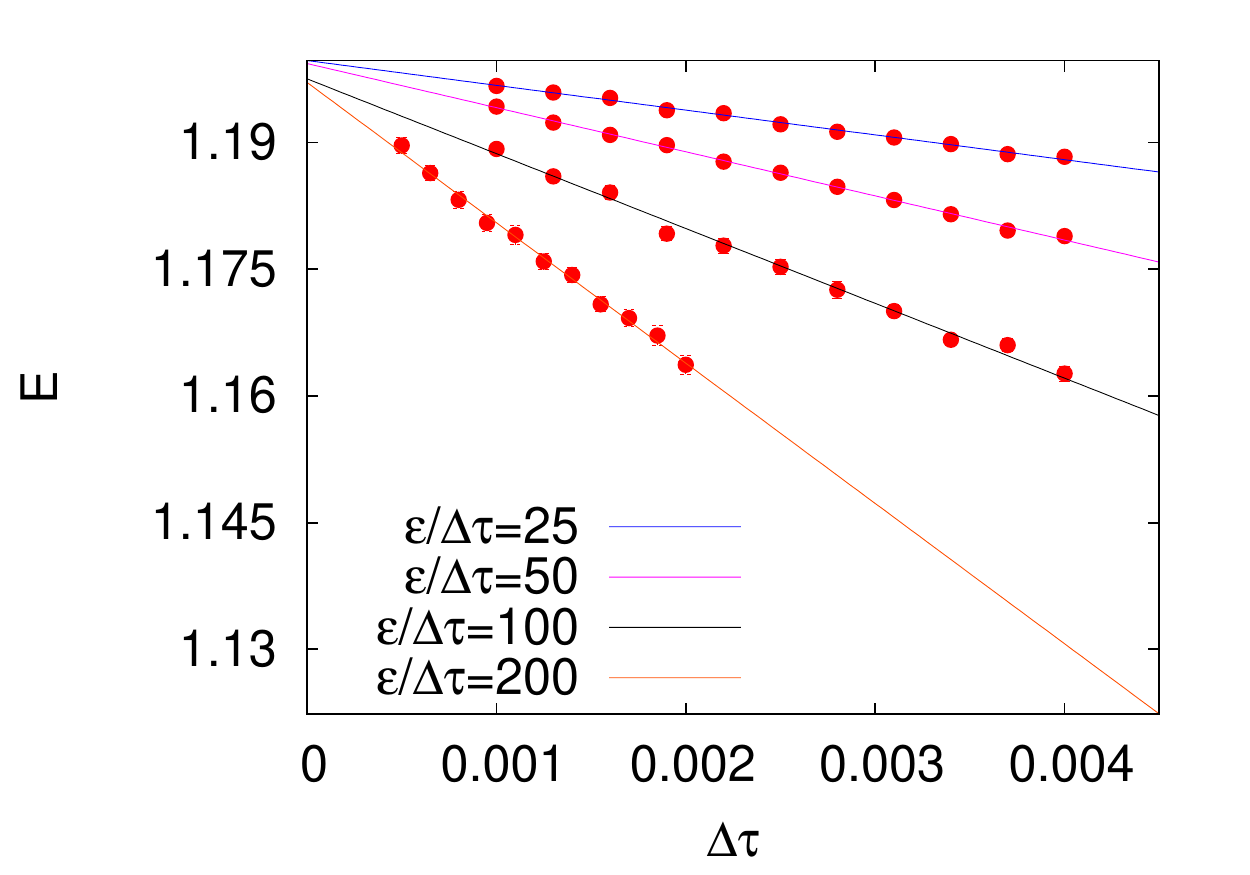}
\caption{Estimation of the DTDMC energy using Method 1 for a one-body
  system with Weyl SOC and a harmonic trap.}
\label{fewbody_dt_eps}
\end{figure}  

\begin{figure}[b]
\centering
\includegraphics[width=0.70\textwidth]{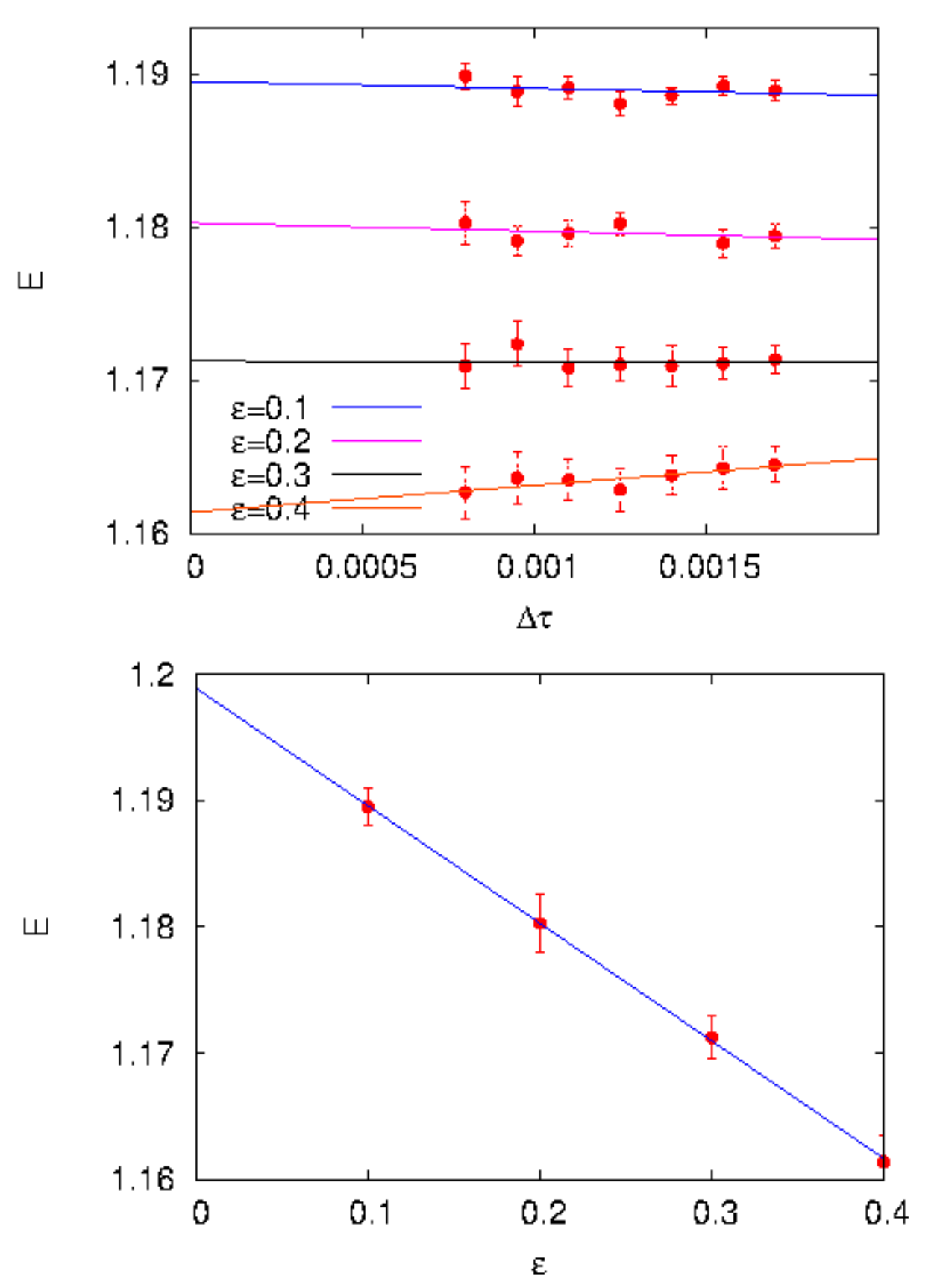}
\caption{Estimation of the DTDMC energy using Method 2 for a one-body
  system with Weyl SOC and a harmonic trap.}
\label{fewbody_fixed_eps_var_dt_comb}
\end{figure}  

\begin{table*}[t]
  \centering
\begin{tabular}{| l | l | l | l | l | l |}
  \hline
   & SIDMC & ITE FPA & DTDMC & DTDMC fixed $\epsilon$ & ITE FPA eff. H \\
  \hline
  Raman & 1.368 $\pm$ 0.001 & 1.3667 $\pm$ 0.0005 & 1.368 $\pm$ 0.001 &  & 1.3679 $\pm$ 0.0005 \\
    \hline
  Weyl & 1.095 $\pm$ 0.002 & 1.0780 $\pm$ 0.0005 & 1.197 $\pm$ 0.002 & 1.190 $\pm$ 0.002 & 1.1887 $\pm$ 0.0005 \\
    \hline
  Rashba 2-b no spin & 1.064 $\pm$ 0.002 & 1.058 $\pm$ 0.003 & 1.148 $\pm$ 0.003 & 1.132 $\pm$ 0.002 & 1.133 $\pm$ 0.003 \\
  \hline 
  Rashba 2-b spin &  &  & 1.279 $\pm$ 0.002 & 1.262 $\pm$ 0.002 & 1.258 $\pm$ 0.003 \\
  \hline 
\end{tabular}
\caption{Results of the energy estimation (in reduced units, see
  Sec. \ref{sec:units}) for the few-body systems described in
  Sec. \ref{sec:fewbody}. Results for the Raman and Weyl cases
  correspond to the total energy while results for the Rashba case
  correspond to the energy per particle.}
\label{tab:table_fewbody}
\end{table*}

In Fig. \ref{fewbody_dt_jsanchez_rashba}, we show the energy as a
function of the imaginary-time step for the two-body Rashba
calculations. We can clearly see a linear dependence of the energy
with the time step, as it corresponds to a linear approximation to the exact propagator. In the DTDMC method, as stated
previously, three limits have to be satisfied in order to obtain the
estimation of the energy: $\Delta \tau \rightarrow 0$, $\epsilon
\rightarrow 0$, and $\frac{\Delta \tau}{\epsilon} \rightarrow 0$. The
extrapolations according to these limits can be performed in several
ways. Here, we present two of them. Method 1 consists on performing
$N_{\text{sets}}$ sets of $N_{\text{sim}}$ simulations making $\Delta
\tau \rightarrow 0$, $\epsilon \rightarrow 0$, with $\frac{\Delta
  \tau}{\epsilon} \ll 1$ fixed. After this, one ends up with
$N_{\text{sets}}$ estimations of the energy, each one associated to a
given $\frac{\Delta \tau}{\epsilon}$ value. Finally, one retains the
estimation associated to the lowest $\frac{\Delta \tau}{\epsilon}$
value. Method 2 consists on performing $N_{\text{sets}}$ sets of
$N_{\text{sim}}$ simulations making $\Delta \tau \rightarrow 0$,
$\frac{\Delta \tau}{\epsilon} \rightarrow 0$, with $\epsilon$
fixed. After this, one ends up with $N_{\text{sets}}$ estimations of
the energy, each one associated to a given $\epsilon$ value. Finally,
one then takes the extrapolation of these estimations in the limit
$\epsilon \rightarrow 0$.

In Figs. \ref{fewbody_dt_eps} and \ref{fewbody_fixed_eps_var_dt_comb},
we show the estimations obtained using Method 1 and Method 2,
respectively, for the one-body system with Weyl SOC and a harmonic
trap. As we can see, the dependence of the energy
extrapolations with respect to $\frac{\Delta \tau}{\epsilon}$ is much
weaker than their dependence on $\epsilon$. Therefore, Method 1 is
preferred and is the one that we have used to provide the T-moves
energy.  We can also see from the figure that the dependence of the
energy with respect to $\Delta \tau$, when $\epsilon$ or $\frac{\Delta
  \tau}{\epsilon}$ are fixed, is linear in both cases. This is because
the non-SOC terms of the propagator are exact up to $\order{\Delta
  \tau}$ while the SOC terms are exact up to $\order{\frac{\Delta
    \tau}{\epsilon}}$. For all the chosen values of $\epsilon$, the
conditions in Eq.~(\ref{condition_eps_exp_2}) are satisfied, with the r.h.s being
$~10^{-2}$. Also, the condition in Eq.~(\ref{condition_eps_el}) is
satisfied since the difference between the r.h.s. and the l.h.s. is at
most a $3 \%$ of the SOC local energy contribution.

From Tab. \ref{tab:table_fewbody}, we can see that both DMC methods
provide energies that agree with the result of the imaginary-time evolution
within a $2\%$ error. We can also see that SIDMC provides lower
energies than DTDMC. This is due to the fixed-phase nature of the
energies obtained with SIDMC, which does not require to use an effective
Hamiltonian as DTDMC. We can see that this effect is enhanced in the
harmonically trapped systems featuring Rashba and Weyl SOCs. For the
cases with two-body spin-dependent interactions, only T-moves results
are reported, since SIDMC can not deal with these kind of
potentials. It must be remarked that, while in the T-moves
calculations we perform the triple extrapolation $\Delta \tau
\rightarrow 0$, $\epsilon \rightarrow 0$, and $\frac{\Delta
  \tau}{\epsilon} \rightarrow 0$, calculations with ITE are performed
at a fixed $\epsilon$
($\epsilon=0.1$ and $\epsilon=0.3$ in the Weyl and Rashba cases, respectively).
This is due to the computational cost of
decreasing $\epsilon$ when discretizing the Schr\"{o}dinger equation
in the position representation, since $\epsilon$ is taken as the
point-to-point distance of the mesh. In order to check that both DTDMC
and ITE give compatible estimates, we also provide in
Tab. \ref{tab:table_fewbody} DMC energies corresponding to a fixed
$\epsilon$. This is not necessary in the Raman case since the Raman
Hamiltonian is independent of $\epsilon$ if this parameter is
sufficiently small, as mentioned previously. Notice also that the
errors corresponding to the ITE results in the two-body 2D Rashba
cases are larger than the ones in the 3D one-body Raman and Weyl
cases. This is due to the higher number of dimensions that must be
discretized in the latter case.

\begin{table*}[t]
  \centering
\begin{tabular}{| l | l | l | l | l | l |}
  \hline
   & SIDMC & GPE FPA & DTDMC & DTDMC fixed $\epsilon$ & GPE FPA eff. H \\
  \hline
  Raman 2-b no spin & -0.0496 $\pm$ 0.0002 & -0.04964 $\pm$ 0.00005 & -0.0496 $\pm$ 0.0004 &  & -0.04962 $\pm$ 0.00005 \\
    \hline
  Raman 2-b spin &  &  & 0.00946 $\pm$ 0.00004 &  & 0.009370 $\pm$ 0.000005 \\
    \hline
  Weyl 2-b no spin & 0.1125 $\pm$ 0.0003 & 0.11217 $\pm$ 0.00005 & 0.1444 $\pm$ 0.0002 & 0.1423 $\pm$ 0.0002 & 0.14239 $\pm$ 0.00005 \\
  \hline     
  Weyl 2-b no spin trial 2 &  &  & 0.1122 $\pm$ 0.00015 & 0.1123 $\pm$ 0.00015 & 0.11225 $\pm$ 0.000005 \\
  \hline 
  Weyl 2-b spin &  &  & 0.0602 $\pm$ 0.0001 & 0.0602 $\pm$ 0.0001 & 0.06029 $\pm$ 0.00005 \\
  \hline 
\end{tabular}
\caption{Results of the energy per particle (in reduced units, see Sec. \ref{sec:units}) for the many-body systems in the dilute regime, as  described in Sec. \ref{sec:dilute}.}
\label{tab:table_dilute}
\end{table*}

\subsection{\label{sec:manybody}Many-body calculations}

We report in this Section the DMC energies corresponding to the many-body Raman and Weyl SOC
Hamiltonians. We first focus on the dilute regime with a finite number of particles imposing periodic boundary conditions
(PBC). We compare the DMC energy estimations with energies obtained
by solving the imaginary time Gross-Pitaevskii equation (GPE), both
for the fixed-phase Hamiltonian (Eq.~(\ref{HFP})) and the fixed-phase,
effective Hamiltonian (Eq.~(\ref{effH_1})).  In the case of Rashba SOC,
we do not know the scattering length of the complete interaction, and thus
a direct comparison to GPE is not possible.  Finally, we compare the
energy estimations of both DMC methods out of the dilute regime.

\subsubsection{\label{sec:dilute}Dilute regime}

Table~\ref{tab:table_dilute} reports the DMC energy per particle
together with the corresponding Gross-Pitaevskii energy per particle,
for four different physical systems: Raman SOC and Weyl SOC, both
with spin-independent and spin-dependent two-body interactions.
Moreover, we include the T-moves energy per particle using two
different trial wave functions in the two-body spin-independent Weyl
case in order to showcase the variational dependence of this method
with respect to the magnitude of the trial wave function.

For the GPE calculations involving Raman or Weyl SOCs, we use the
free-space scattering length, i.e., the scattering length obtained for
the Hamiltonian removing the SOC terms~\cite{raman_scattering,
 weyl_scattering}.

In all cases, the trial wave function is of the form
\begin{align}
 &\Psi_T(\vec{R},\vec{S}) = \left[ \prod_{j=1}^N \rho_{T,\text{1b}} (\vec{r}_j,s_j) \right] \prod_{ \substack{i,j=1 \\ i<j} }^N \overline{\rho}_{T,\text{2b}} (\vec{r}_i,\vec{r}_j) \nonumber \\
 & \cross \exp \left[ i \sum_{j=1}^2 \phi_T (\vec{r}_j,s_j) \right] \ ,
 \label{twf_mb_general}
 \end{align}
with
\begin{align}
 &\overline{\rho}_{T,\text{2b}} (r_{ij}) = \begin{cases}
                                            \frac{ \rho_{T,\text{2b}} (r_{ij}) + \rho_{T,\text{2b}} (L - r_{ij}) }{ 2 \rho_{T,\text{2b}} (L/2) } & \text{ if } r_{ij} < L/2
                                            \\
                                            1 & \text{ if } r_{ij} > L/2
                                           \end{cases} \label{rho2bpbc_new}
\end{align}
and $r_{ij} = \abs{\vec{r}_j-\vec{r}_j}$.  The function
$\rho_{T,\text{2b}} (r_{ij})$ is the magnitude of a spin-independent two-body trial
wave function analogous to the one presented in Sec. \ref{sec:fewbody}
(here $k_{\text{2b}} \sim 10^{-6}$). The magnitude of the one-body terms for the
T-moves "Raman 2-b no spin" and "Raman 2-b spin" cases are given in
Eqs.~(\ref{raman_mod_up}) and~(\ref{raman_mod_down}). For the SIDMC "Raman 2-b
no spin" case we use the expression in
Eq.~(\ref{raman_mod}). Both DTDMC and SIDMC "Weyl 2-b no spin" cases are done with the terms in Eq.~(\ref{weyl_mod}), while in the T-moves
"Weyl 2-b no spin trial 2" and "Weyl 2-b spin" cases we use the
one-body forms of Eqs.~(\ref{weyl_mod_up}) and~(\ref{weyl_mod_down}). In all cases no harmonic trap has
been used.
The trial phases for each case are analogous to the ones in
Eqs.~(\ref{raman_phase_up}),~(\ref{raman_phase_down}),~(\ref{weyl_phase_up}), and~(\ref{weyl_phase_down}).

The average number of walkers is set to $N_w = 1,000$ and the
time step is $\Delta \tau \sim \order{10^{-3}}$. The parameter
$\epsilon$ of DTDMC is fixed as $\epsilon = 100 \Delta \tau$. All the
used values of $\epsilon$ satisfy the condition of
Eq.~(\ref{condition_eps_el}), with a discrepancy between the r.h.s. and
the l.h.s. of at most $\sim 1 \%$.  Also, the r.h.s of both expressions in
Eq.~(\ref{condition_eps_exp_2}) equals
$0.08$ at most, which implies that the maximum error in the
approximation to the propagator is $e_{\text{max}} \sim e^{0.08} - (1
+ 0.08) \simeq 0.0033$. In the Weyl SIDMC calculations, the length of a simulation
block is set to $N_b = 10$. The ratio of eliminated
walkers is $\chi < 0.0002$. For the Raman calculations, we have
$N_b = 10$ and $\chi = 0$ (see Sec. \ref{sec:algorithm}).

The Raman simulations are carried out with $N=40$ particles,
$\eta_{\text{Rm}} = 0.4$, $L_x = L_y = L_z = 16.899$ (box length) and
$k = k_x = \frac{2 \pi}{L_x}$. In the two-body spin-independent case
we have $V_0 = 75$, $R_0 = 0.25$, $\Omega = 0.4$, $C_1 =0$, $C_2 =1$
and $B_c = 0.5$, while in the two-body spin-dependent case we have
$V_0(+1,+1) = V_0(-1,-1) = 75$, $V_0(+1,-1) = V_0(-1,+1) = 50$, $R_0 =
0.25$, $\Omega = 0.1$, $C_1 = 0.6$, $C_2 = 0.8$. The gas parameter for
these systems is $n a^3 \simeq 10^{-6} $.

In the Weyl simulations, and for the two-body spin-independent case, we
use $N=45$ particles, $\eta_{\text{We}} = 0.25$, $L_x = L_y = L_z =
20$, $\vec{k} = (k_x,0,k_z)$ with $k_i = \frac{2 \pi}{L_i}$, $V_0 =
75$, $R_0 = 0.3$, with a gas parameter of $na^3 = 1.7 \times
10^{-5}$. In the two-body spin-dependent case we use $N=35$,
$\eta_{\text{We}} = 0.25$, $L_x = L_y = L_z = 18$, $k = k_x = \frac{2
  \pi}{L_x}$, $V_0(+1,+1) = V_0(-1,-1) = 75$, $V_0(+1,-1) = V_0(-1,+1)
= 50$, $R_0 = 0.3$, with a gas parameter of $na^3 \sim 10^{-5}$.

We can see from Table~\ref{tab:table_dilute} that the DMC energies
agree with the GPE calculations up to a $\sim 1 \%$. As in the
previous Section, for the spin-dependent two-body cases only T-moves
results are reported, since the SIDMC method can not solve two-body
spin-dependent interactions. We can also see from the two-body
spin-independent cases that DTDMC is able to recover almost
completely the fixed-phase energy, although we know it always provides an upper bound 
to it.
On the other hand, SIDMC
recovers the complete fixed-phase energy. 
The DTDMC Weyl two-body spin-independent calculations illustrate the variational property with respect to the magnitude of the trial wave function of this method. Notice that two different magnitudes ("Weyl 2-b no spin" and "Weyl 2-b no spin" cases) provide two different energy estimations. 

\begin{table*}[t]
  \centering
\begin{tabular}{| l | l | l |}
  \hline
   & SIDMC & DTDMC \\
  \hline
  Raman PBC 2-b no spin & 3.673 $\pm$ 0.002 & 3.681 $\pm$ 0.002 \\
    \hline
  Raman PBC 2-b spin trial 1 &  & 5.356 $\pm$ 0.003 \\
    \hline
  Raman PBC 2-b spin trial 2 &  & 5.358 $\pm$ 0.002 \\
  \hline 
  Weyl PBC 2-b no spin & 3.773 $\pm$ 0.003 & 3.798 $\pm$ 0.003 \\
  \hline 
  Weyl PBC 2-b no spin trial 2 &  & 4.050 $\pm$ 0.005 \\
  \hline 
  Weyl PBC 2-b spin &  & 5.633 $\pm$ 0.005 \\
  \hline 
  Weyl HO 2-b no spin & 2.236 $\pm$ 0.001 & 2.302 $\pm$ 0.002 \\
  \hline 
\end{tabular}
\caption{Energies (in reduced units, see Sec. \ref{sec:units}) for the
  many-body systems out of the dilute regime, as described in
  Sec. \ref{sec:out_of_dilute}.}
\label{tab:table_out_of_dilute}
\end{table*}

\subsubsection{\label{sec:out_of_dilute} Beyond the dilute regime}

In this Section we compare the performance of the two DMC algorithms
discussed in several homogeneous many-body systems, beyond the dilute
regime. We analyze a few systems featuring Raman and Weyl SOCs using
periodic boundary conditions, and a two-body spin-independent
interaction. We show again an example of the variation of the T-moves
energy when the magnitude of the trial wave function is changed. We also provide DTDMC
energy estimations of systems under Raman and Weyl SOCs with a
spin-dependent two-body interaction. Finally, we compare both DMC
estimations in a many-body harmonically confined system with Weyl SOC. Results are presented in
Table~\ref{tab:table_out_of_dilute}.

The general form of the trial wave function is given in
Eq.~(\ref{twf_mb_general}). The T-moves calculations corresponding to
the cases "Raman PBC 2-b no spin", "Raman PBC 2-b spin trial 1" and
"Raman PBC 2-b spin trial 2" use the one-body terms of
Eqs.~(\ref{raman_mod_up}) and~(\ref{raman_mod_down}), while for the SIDMC
"Raman PBC 2-b no spin" calculation Eq.~(\ref{raman_mod}) has been used. For DTDMC corresponding to the cases "Weyl
PBC 2-b no spin", "Weyl PBC 2-b spin", and "Weyl HO 2-b no spin" we
use the expressions in Eqs.~(\ref{weyl_mod_up}) and~(\ref{weyl_mod_down})
while for the DTDMC "Weyl PBC 2-b no spin trial 2" case we use

\begin{figure}[t]
\centering
\includegraphics[width=0.85\linewidth]{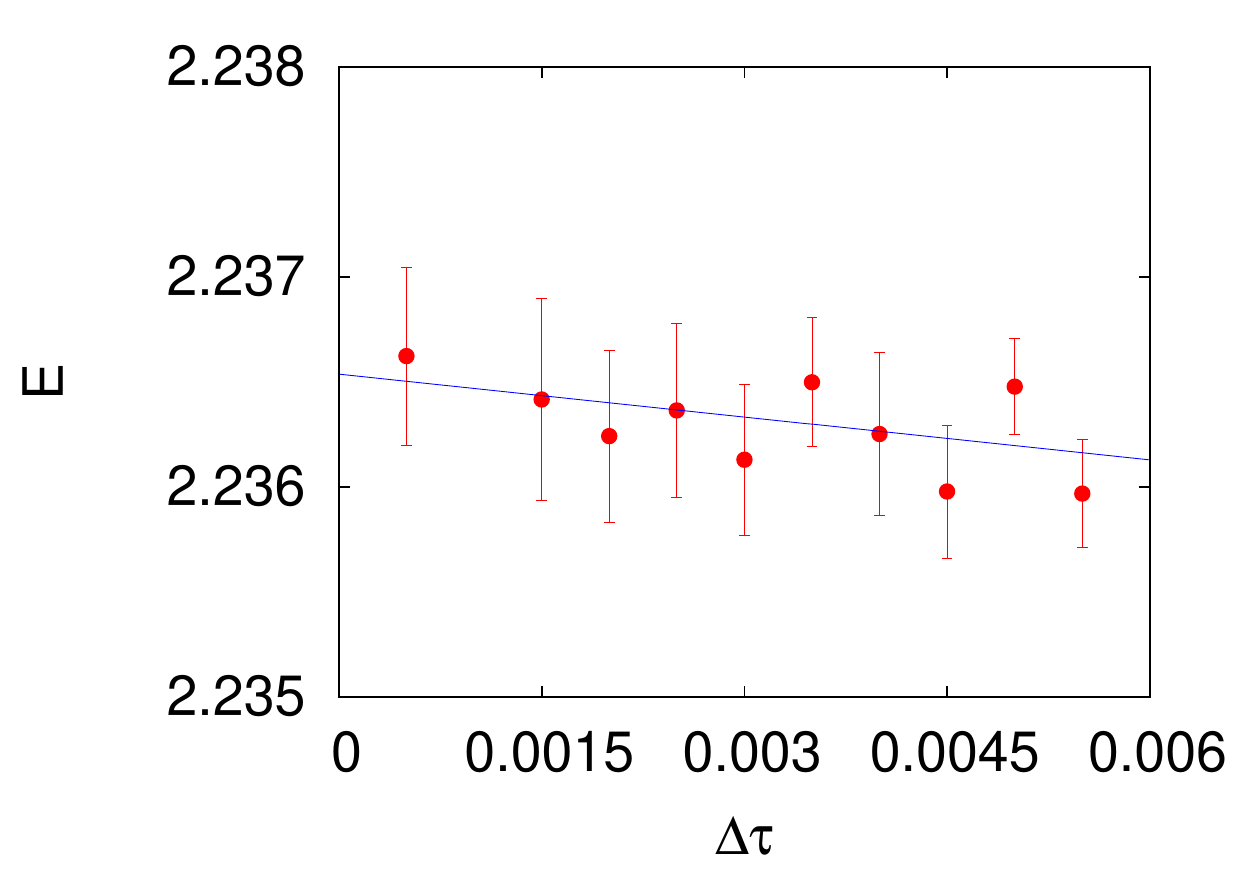}
\caption{Dependence of the DMC energy on the imaginary time-step for  SIDMC for a many-body system with Weyl SOC and a harmonic trap.}
\label{manybody_dt_jsanchez}
\end{figure}  

\begin{figure}[t]
\centering
\includegraphics[width=0.85\linewidth]{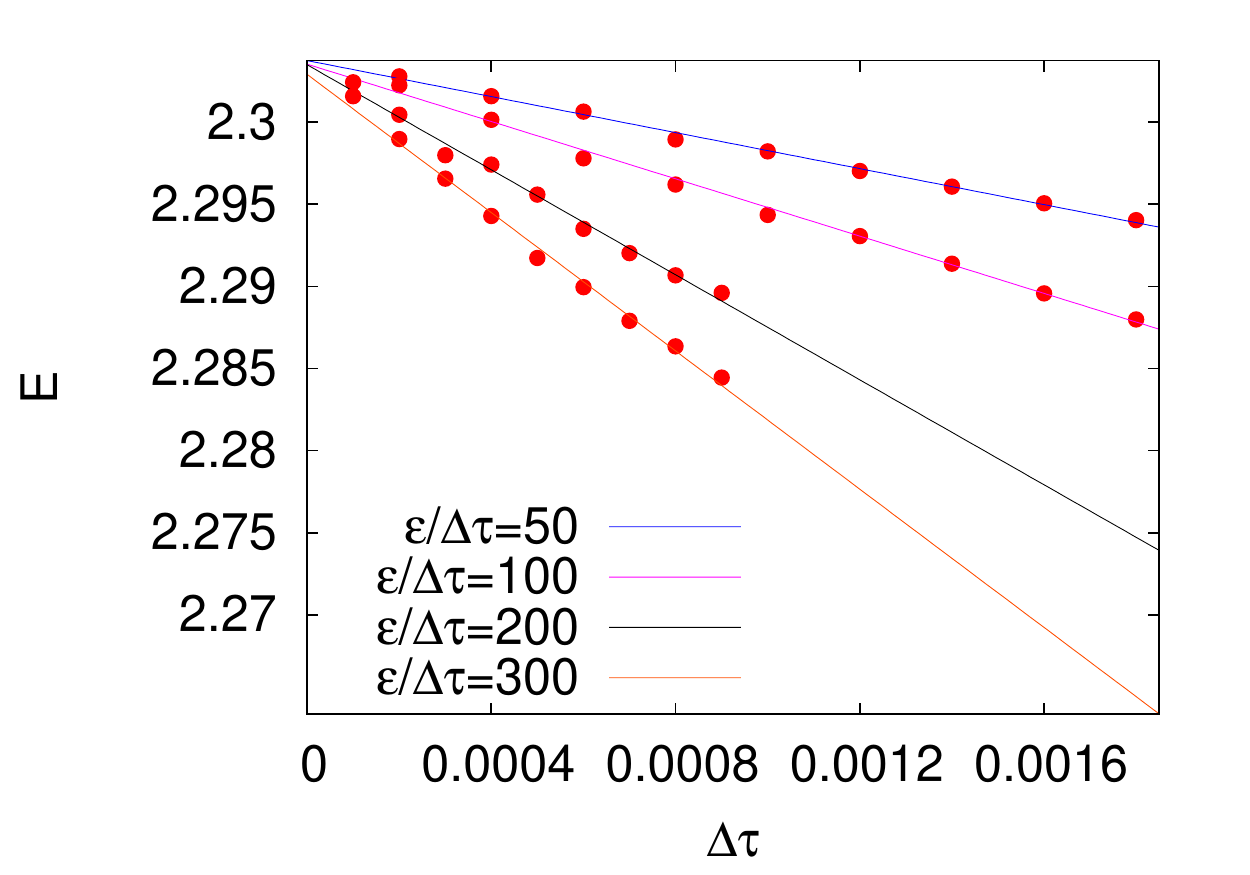}
\caption{Estimation of the DTDMC energy using Method 1 for a many-body system with Weyl SOC and a harmonic trap.}
\label{manybody_dt_eps_1}
\end{figure}  

\begin{figure}[b]
\centering
\includegraphics[width=0.85\textwidth]{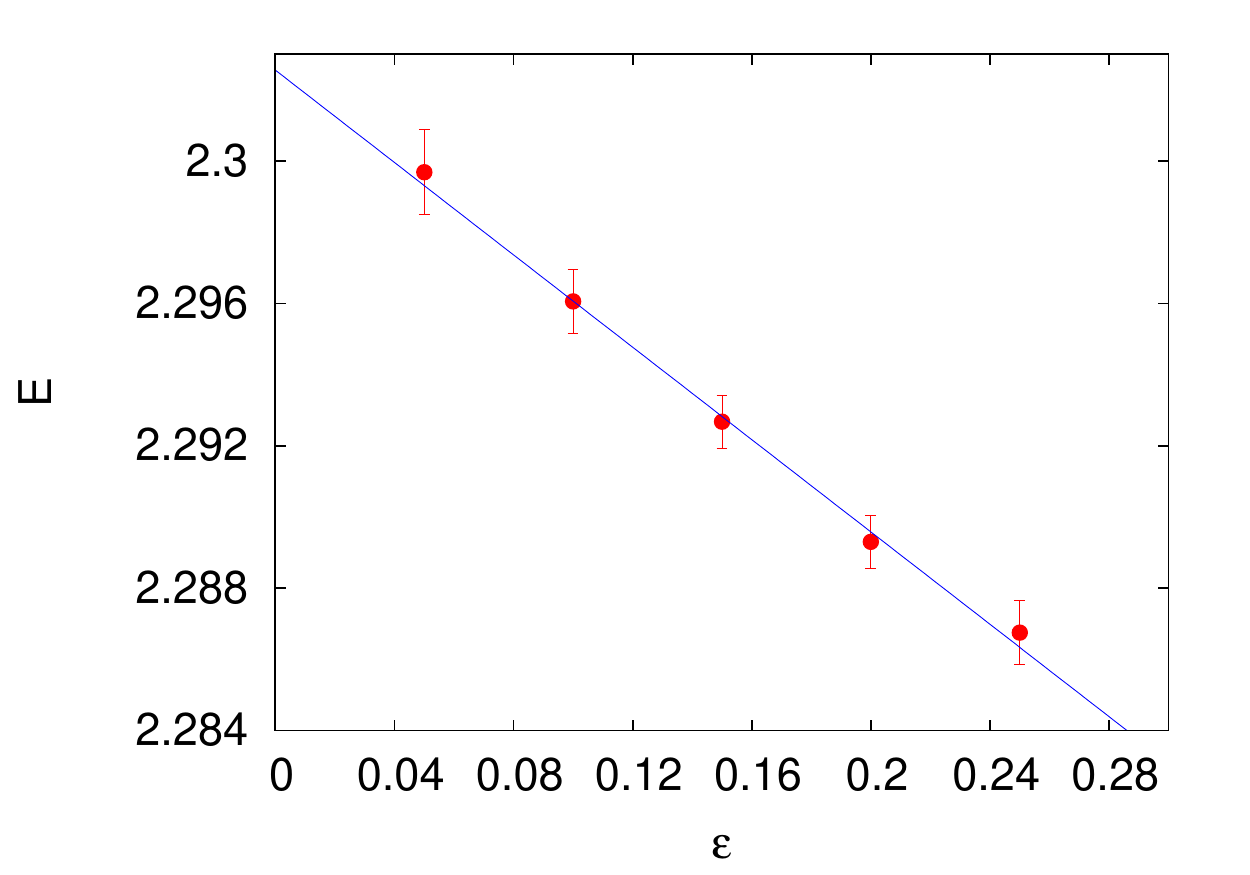}
\caption{Estimation of the DTDMC energy using Method 2 for a many-body system with Weyl SOC and a harmonic trap.}
\label{manybody_fixed_eps_var_dt_1}
\end{figure}  

\begin{align}
 &\rho_T(\vec{r},s=+1) = \gamma
 \label{weyl_mod_up_worse}
 \\
 &\rho_T(\vec{r},s=-1) = \sqrt{1 - \gamma^2} \frac{ \left( 1 + \cos \theta_k \right) }{ \sin \theta_k } 
 \label{weyl_mod_down_worse}
 \\
 &\gamma = 0.6
\end{align}
This form helps us to illustrate the variational property
of the T-moves method with respect to the magnitude of the trial wave function. The SIDMC
"Weyl PBC 2-b no spin" and "Weyl HO 2-b no spin" calculations use the
expressions in Eq.~(\ref{weyl_mod}). As in the previous Section, the
trial phases for each case are 
given in
Eqs.~(\ref{raman_phase_up}),~(\ref{raman_phase_down}),~(\ref{weyl_phase_up}), and~(\ref{weyl_phase_down}).

In the two-body spin-independent calculations, the two-body trial
terms in all PBC cases are the same as in
Sec. \ref{sec:dilute}. Concerning the two-body spin-dependent
calculations, we report the energy in the Weyl case using a
spin-independent two-body correlation factor analogous to the one in
Sec. \ref{sec:dilute}. In the Raman case, though, we compare the energy
estimated using a spin-independent two-body factor with that
estimated using a spin-dependent one, 
again with the same form as in Sec. \ref{sec:dilute}. Finally, in the "Weyl HO 2-body no spin" 
case we set $\overline{\rho}_{T,\text{2b}} (r_{ij}) = \rho_{T,\text{2b}} (r_{ij})$ 
in Eq.~(\ref{rho2bpbc_new}) because we do not impose PBC.

The average number of walkers is set to $N_w = 1000$, the time step
$\Delta \tau \in ( 10^{-4}, 10^{-3} )$, and the DTDMC $\epsilon$ parameter
is fixed such that $\frac{\epsilon}{\Delta \tau} \in (100,
400)$ for Weyl and $\frac{\epsilon}{\Delta \tau} = 10$ for Raman. All
the used values of $\epsilon$ satisfy the condition in
Eq.~(\ref{condition_eps_el}), with a discrepancy between the r.h.s. and
the l.h.s. of at most $3 \%$. Also, the r.h.s of both expressions in
Eq.~(\ref{condition_eps_exp_2}) equals
$0.3$ at most, which implies that the maximum error in the
approximation to the propagator is $e_{\text{max}} \sim e^{0.3} - (1 +
0.3) \simeq 0.05$. In the Weyl PBC SIDMC calculations the length of a simulation
block is set to $N_b = 10$. The ratio of eliminated
walkers is $\chi < 0.006$. The harmonically trapped Weyl simulations
share the same parameters except for the ratio of eliminated walkers, $\chi < 0.001$. For the Raman calculations one has
$N_b = 10$ and $\chi = 0$ (see Sec. \ref{sec:algorithm}).

In the Raman case we use $N=50$ particles, $\eta_{\text{Rm}} = 1.5$,
$\Omega = 0.4$, $L_x = L_y = L_z = 4.5$, $V_0 = 1$, $R_0 = 1.5$, $k =
\frac{2 \pi}{L_x}$, and $C_1 = 0.6$, $C_2 = 0.8$. In the SIDMC
simulations we also have $B_c = 0.5$. The two-body spin-dependent case
shares the same parameters with the exception of $V_0(+1,+1) =
V_0(-1,-1) = 2$, $V_0(+1,-1) = V_0(-1,+1) = 1$. The gas parameter for
the up-down channels is $n a^3 \sim 10^{-2}$ while for the up-up and
down-down channels we set $n a^3 \sim 0.1$. In the PBC two-body
spin-independent Weyl case we simulate $N=25$ particles with
$\eta_{\text{We}} = 3.590$, $L_x = L_y = L_z = 3.5$, $V_0 = 1$, $R_0 =
1.5$, and $\vec{k} = (k_x,0,0)$ $k_x = \frac{2 \pi}{L_x}$. The
two-body spin-dependent case shares the same parameters with the
exception of $V_0(+1,+1) = V_0(-1,-1) = 2$, $V_0(+1,-1) = V_0(-1,+1) =
1$. The gas parameter for each channel is of the same order of
magnitude that the one in the Raman case. Finally, in the harmonically
trapped Weyl simulations we use $N=30$ particles, $\eta_{\text{We}} =
1$, $\omega = 0.4$, $V_0 = 1$, $R_0 = 1.5$, $k = 0.5$, $\theta_k =
1.31$, and $\phi_k = 0.3$.

In Fig. \ref{manybody_dt_jsanchez}, we show the energy dependence on
the imaginary time-step corresponding to the SIDMC simulations of trapped
Weyl gases. We can see in the Figure the linear dependence of
the energy with respect to $\Delta \tau$.  In
Figs. \ref{manybody_dt_eps_1} and \ref{manybody_fixed_eps_var_dt_1},
we show DTDMC results for the two methods mentioned in
Sec. \ref{sec:fewbody} to estimate the triple limit $\Delta \tau
\rightarrow 0$, $\epsilon \rightarrow 0$, and $\frac{\Delta
  \tau}{\epsilon} \rightarrow 0$. The observed behavior is consistent
with the previous results obtained in the one-body case.

In Table \ref{tab:table_out_of_dilute}, we report the DMC energies for
the analyzed cases.  From these results, we can see that DTDMC is able to almost exactly recover the fixed-phase energy of the bulk gases.  In
the trapped Weyl gas, the difference with respect to the fixed-phase
energy obtained with SIDMC is larger. We can also see how the
improvement of the magnitude of the trial wave function in the two-body spin-independent
PBC Weyl simulation produces better energies as a consequence of the
variational property of the DTDMC method. Finally,
our results show that the spin-dependent two-body trial correlation factor does
not make any significant difference in the two-body spin-dependent PBC
Raman simulation.

\section{\label{sec:conclusions}Conclusions}

In this paper, we discuss two different Diffusion Monte Carlo methods (DTDMC and SIDMC) that are able to deal with many-body
systems of ultracold quantum gases featuring synthetic Spin-Orbit
Coupling. DTDMC
is an extended  version of the method of
Refs.~\cite{mitas} and \cite{tmoves} to the relevant SOC interactions in the field of ultracold gases, but with discrete spins. This method relies on the
introduction of an effective Hamiltonian and provides an upper bound to the fixed-phase
energy of the system. On the contrary, the SIDMC method is able to avoid this
issue by propagating the spin-integrated probability density,
providing exact fixed-phase estimations. However, SIDMC is not
able to deal with spin-dependent two-body interactions and requires the use of
spin-independent trial wave functions.

We have described the formalism of both methods in detail, together
with a scheme of both algorithms for future applications.  We have
reported the energy estimation of several few-body systems, featuring
three different kinds of SOC interactions. We have compared these results with
energies obtained by propagating the Schr\"{o}dinger equation in
imaginary time, finding good agreement between both estimations. We
have also performed simulations of many-body systems in the dilute regime and
have recovered the energies obtained by solving the imaginary time
Gross-Pitaevskii equation with discrepancies of at most $\sim 1
\%$. Finally, we have compared both algorithms beyond the dilute
regime, showing that the DTDMC method is able to recover the
fixed-phase energy almost completely in the PBC cases. We hope that
these methods can be used to explore the physics of SOC systems beyond the mean field, dilute regime.

\begin{acknowledgments}
We acknowledge partial financial support from MINECO Grants No. FIS2014-56257-C2-1-P and No. FIS2017-84114-C2-1-P. J. S\'anchez-Baena also acknowledges the FPU fellowship with reference FPU15/01805 from MECD.

\end{acknowledgments}

\newpage

\setcounter{section}{0}

\end{document}